\documentclass[12pt]{iopart}

\usepackage{hyperref}
\usepackage{graphicx}
\usepackage{rotating}
\usepackage{array}
\usepackage{amssymb}

\def\v#1{{\bf#1}}
\def\be{\begin{equation}}
\def\ee{\end{equation}}
\def\bea{\begin{eqnarray}}
\def\eea{\end{eqnarray}}
\def\ahalf{{\textstyle{1\over2}}}

\newcommand{\bfalpha}{\mbox{\boldmath$\alpha$\unboldmath}}

\newcommand{\bftau}{\mbox{\boldmath$\tau$\unboldmath}}

\def\ie{{\it i.e.\,}}
\def\etal{{\it et al.   }}

\def\kcal{\mbox{$\cal K\,$}}
\def\<{\langle}
\def\>{\rangle}

\begin{document}

\title{Propagators in two-dimensional lattices}
\author{E. Sadurn\'i }

\address{Instituto de F\'isica, Benem\'erita Universidad Aut\'onoma de Puebla,
Apartado Postal J-48, 72570 Puebla, M\'exico}

\eads{ \mailto{sadurni@ifuap.buap.mx}}

\begin{abstract}

\noindent
This paper is devoted to the computation of discrete propagators in two-dimensional crystals and their application to a number of time dependent problems. The methods to compute such kernels are provided by a tight-binding representation of Dirac matrices and the generalizations of Bessel functions. Diffusive effects of point-like distributions on crystalline sheets are studied in a second quantization scheme. In the last part, a compendium of propagators is presented. The cases of square, triangular and hexagonal arrays are covered.

\end{abstract}


\maketitle

\section{Introduction}

It is well known that quantum mechanics in discrete variables, tight-binding models in solids and electromagnetic waves in periodic media share similar dynamical formalisms. In the traditional approach to these systems, the band structure plays a central role in the determination of the propagating properties of the medium. These properties are usually inferred from the corresponding dispersion relations, which may contain spectral gaps and conical points. Such a traditional approach to the study of wave propagation has been applied extensively in solid state physics and photonic crystals. Simplified models of graphene \cite{semenoff, wallace} and their mesoscopic emulations \cite{sadurni, richter} lie in these categories.

It is rarely seen, however, a detailed description of the propagation of wavepackets in crystals as a function of time. In the case of solid state physics, we might conjecture that the lack of such theoretical studies obeys limited experimental accessibility, as there is no apparent need for careful theoretical predictions. The Fermi velocity of electrons in a good conductor ($\sim 10^6$m$/$s \cite{Ash}) is a signature of their high mobility and makes the detection of quantum transient effects far from trivial. A similar situation seems to hold for electromagnetic waves in dielectrics arranged in crystalline structures, in view of a high speed of propagation.

Despite the previous observations, the possibility of studying the time behavior of waves in a `slow' crystalline medium is not technically precluded. For example, the detection of vibrations as a function of time in a metallic rod or plate with carved periodic structures should be possible within the state of the art of acoustics \cite{flores, flores2, flores3}, with typical speeds of propagation $\sim 10^3$m$/$s. For quantum-mechanical waves, the detection of time-dependent effects might be feasible as well: matter waves in the form of Bose-Einstein condensates are now being produced and trapped in optical lattices emulating crystals \cite{spielman, cheneau, ibloch}.
 
The necessity of time-dependent descriptions due to these realizations is nowadays more palpable.
Moreover, even if the evolution of wavepackets representing electrons in solids or two-dimensional materials (e.g. graphene) might be difficult to observe, we would expect that the time-dependent effects appearing in slow realizations should also take place in their faster counterparts.

It is worthwhile to mention that in the context of graphene, time-dependent descriptions of wavepackets are available in the literature \cite{kramer}. The effective Dirac description of electrons in the honeycomb lattice pays off, as the Dirac propagator can be written explicitly \cite{berry, sym}. Interesting effects such as zitterbewegung have been predicted \cite{rusin}. Nevertheless, we recognize that such treatments are valid only for low energies (e.g. near conical points), ignoring the effects due to the cut-off energy and the periodicity of the medium. As a result, these approximations restrict the types of wavepackets that one may study: well-localized distributions do not fit into this scheme due to large uncertainties in the momenta.

\pagebreak

For the reasons above, in this paper we develop the formalism of propagators for quantum-mechanical waves in two-dimensional {\it static\ }crystals regarded as tight-binding lattices, providing thus a natural continuation to our previous works \cite{5, 5.1}. Our formalism allows the computation of propagators corresponding to the three possible types of periodic {\it regular\ }lattices in two-dimensions: square, triangular and hexagonal tessellations. As an immediate application we describe the diffusive process in time of a wavepacket initially located at a single site. This may represent an electron initially found in an atomic state or $-$if we confine our attention to a component of the electromagnetic field$-$ a light pulse originally fed to a resonant site in a photonic structure. We also provide the corresponding transition amplitudes when particle statistics are incorporated in a second quantization scheme, giving the opportunity to discuss simplified few-body problems. 

The paper is organized as follows: In the next section, we establish the notation and the formalism to be used throughout the paper. In subsection \ref{s2.1} we discuss Green's functions defined on lattices rather than in continuous space; this is done for the sake of clarity. In subsection \ref{s2.2} we present a second quantization scheme in general lattices and in subsection \ref{s2.3} we give the transition amplitudes for multiparticle states in terms of propagators; detailed derivations are left to the appendices. Section \ref{s3} is devoted to the mathematical tools for the evaluation of kernels. These include a superalgebra $S(2)$ containing a Clifford algebra for bipartite crystals, a generalization of the Bessel function for triangular lattices and a number of Bessel expansions that appear in connection with spectral gaps. One of such expansions allows to compute the propagator of electrons in a model of graphene, section \ref{graphene}. In section \ref{s4} we study the diffusion of point-like distributions and the dynamics of identical particles in simplified examples. In section \ref{s5} a space-time approach to the Dirac limit in hexagonal and linear lattices is presented, providing a continuous limit of the theory in terms of couplings. Section \ref{summary} is a compendium of results for various lattices. We conclude in section \ref{s7}.

\section{Discrete propagators: General considerations}

We start by fixing some concepts and notation. In tight-binding arrays on a crystalline structure we can consider on-site (or atomic) wavefunctions $\xi_{\v A}(\v x)=\< \v x | \v A \>$ with $\v x$ the position variable in continuous space and $\v A$ a vector which is a member of the underlying lattice. A hamiltonian operator with nearest-neighbor interactions on a homogeneous infinite array can be written as 

\bea
H = \Delta \sum_{\v A, i} | \v A  \> \< \v A + \bfalpha_i | + \mbox{h.c.} + E_0 \sum_{\v A} | \v A  \> \< \v A | 
\label{I1}
\eea
where $\bfalpha_i$ are vectors connecting a site with all nearest neighbors, $i=1,...,c$ and $c$ is the coordination number. Occasionally, we shall be interested in two atomic species described by sublattices A and B with vectors $\v A$ and $\v B$ respectively. This leads to hamiltonians of the type

\bea
H = \Delta \sum_{\v A, i} | \v A  \> \< \v A + \bfalpha_i | + \mbox{h.c.} + E_1 \sum_{\v A} | \v A  \> \< \v A | + E_2 \sum_{\v B} | \v B  \> \< \v B | 
\label{I2}
\eea
and such that $\v A + \bfalpha_i$ is a vector of the sublattice B. Hamiltonians of this type can be constructed for hexagonal lattices (e.g. graphene, boron nitride), for square lattices when viewed as two interpenetrating square lattices of twice the period and even for one-dimensional chains of dimers constructed from two monomeric (homogeneous) chains. Specific notation with primitive vectors for the various types of lattices can be found in sections \ref{square}, \ref{triangle} and  \ref{hexagon}. For one-dimensional chains, lattice A stands for points parameterized by even integers and B for points parameterized by odd integers.

\begin{figure}[!h] \begin{center} \begin{tabular}{ccc} \includegraphics[scale=0.25]{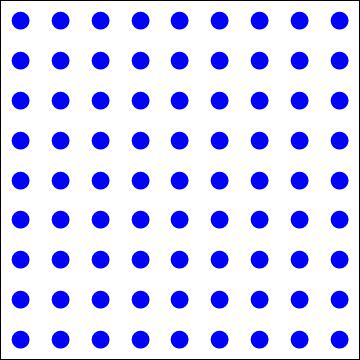} &  \includegraphics[scale=0.25]{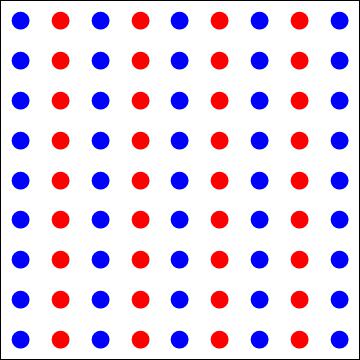} &  \includegraphics[scale=0.25]{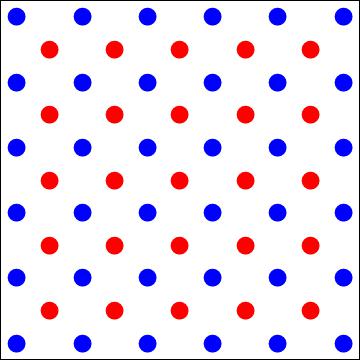} \\   \includegraphics[scale=0.25]{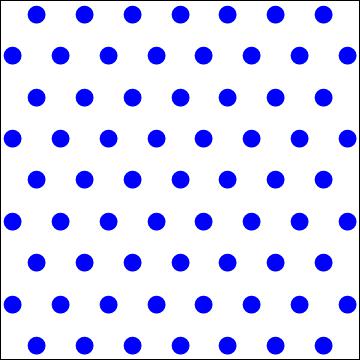} &  \includegraphics[scale=0.25]{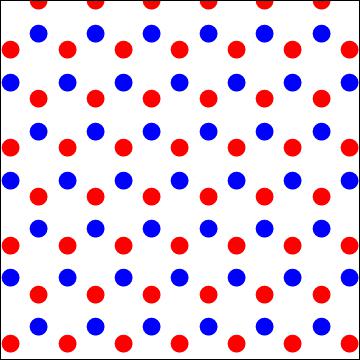} &  \includegraphics[scale=0.25]{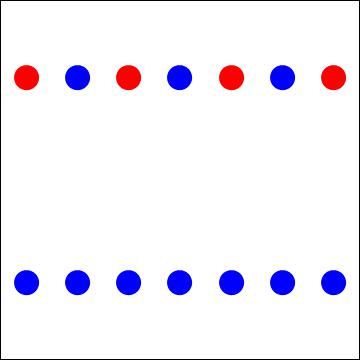} \end{tabular} \end{center} 
\caption{ \label{fig:lattices}Types of lattices considered in this paper. We concentrate on regular arrays of one and two species. The last panel shows two types of linear chains that shall be considered as well: monomeric (blue) and dimeric (red-blue) arrays. Their propagators can be used in the construction of kernels for two dimensional structures.} \end{figure}

No other {\it regular\ }lattices are possible, except for those in which the interactions extend beyond nearest neighbors or those in which several atomic species are considered. See figure \ref{fig:lattices}.

\subsection{Discrete propagators as Green's functions} \label{s2.1}

The fact that a propagator can be identified with the time-dependent Green's function of the Schr\"odinger operator is related to a pair of properties worthwhile to recall at this point: 1) The spectral decomposition of the evolution operator and 2) the boundary condition at the origin of time, precluding any possibility of {\it backward\ } propagation. We expect similar properties in our discrete settings, with adequate normalization constants obtained by summation rather than integration (Dirac deltas in discrete space are not allowed). Let us start from our basic principles. We denote the reciprocal lattice vectors by $\v k$ and the Bloch states by $|\phi(\v k) \>$. The Bloch states are given by superpositions of atomic states in the form

\bea
| \phi(\v k) \> = \frac{1}{\sqrt{\omega}} \sum_{\v A} e^{i \v k \cdot \v A} | \v A \>
\label{II.2}
\eea
where $\omega$ is the volume of the first Brillouin zone $\Omega_B$. These states satisfy

\bea
 H | \phi(\v k) \> = E_{\v k} | \phi(\v k) \>
\label{II.3}
\eea
with $H$ as in (\ref{I1}) and
\bea
\int_{\Omega_B} d\v k \< \phi(\v k)  | \phi(\v k')  \> = \delta( \v k - \v k').
\label{II.4}
\eea
Choosing units in which $\hbar=1$, the equation of the evolution operator reads

\bea
\left[ H -  i \frac{\partial}{\partial t}\right] U_{t,t'} = 0
\label{II.1}
\eea
and with (\ref{II.3}), we can have a spectral decomposition of $U$ as follows

\bea
U_{t,t'} = \int_{\Omega_B} d\v k  | \phi(\v k) \>\< \v \phi(\v k) | e^{-i E_{\v k} (t-t')} .
\label{II.5}
\eea
The propagator $K$ can be defined by taking the matrix elements of (\ref{II.5}) between atomic states and imposing the boundary condition of forward propagation, namely

\bea
K(\v A, \v A'; t,t') = \theta(t-t') \< \v A | U_{t,t'}| \v A' \>.
\label{II.6}
\eea
In order to understand the role of $K$, we need a shorthand for the action of $H$ on functions $f$ of lattice points $\v A$, \ie 
\bea
\hat H f(\v A) \equiv \sum_{\v A' } \< \v A | H | \v A' \> f(\v A').
\label{II.6.1}
\eea
 The differential representation of $\hat H$ can be obtained from (\ref{I1}) by recognizing the presence of translation oparators $T_i=\exp (i \v p \cdot \bfalpha_i)$, \ie

\bea
\hat H = E_0 \v 1 + \Delta \sum_{i=1}^{c}\left( T_i + T_i^{\dagger} \right).
\label{II.10}
\eea
Thus, for arbitrary $f$ we write

\bea
\hat H f(\v A) =  E_0 f(\v A) + \Delta \sum_{i=1}^{c} \left[ f(\v A + \bfalpha_i) + f(\v A - \bfalpha_i) \right].
\label{action}
\eea
Finally, we can apply the rule (\ref{action}) to $K$ and find that it corresponds to the Green's function of the Schr\"odinger operator: Using (\ref{II.6}) and the evolution equation (\ref{II.1}), we find 
\bea
\left[ \hat H -  i \frac{\partial}{\partial t}\right] K(\v A,\v A';t) = -i \delta(t) \delta_{\v A,\v A'}
\label{II.7}
\eea
where $\delta_{\v A,\v A'}$ is a Kronecker delta of lattice points. Additionally, the energy-dependent Green's function $\kcal_E$ can be obtained by a semi-Fourier transform of $K$, which gives

\bea
\kcal_E(\v A, \v A') = i \int_{\Omega_B} d\v k \frac{\< \v A | \phi(\v k) \>\< \v \phi(\v k) | \v A' \>}{E-E_{\v k}} = \frac{i}{\omega}  \int_{\Omega_B} d\v k \frac{e^{i \v k \cdot (\v A - \v A')}}{E-E_{\v k}}
\label{II.8}
\eea
and it satisfies the usual equation
\bea
\left[ \hat H - E \right]\kcal_E(\v A, \v A') = - i \delta_{\v A, \v A'}.
\label{II.9}
\eea
It will be customary to parameterize our results with sets of integers. Let us express the lattice points $\v A, \v A'$ in terms of primitive vectors $\v a_1, \v a_2$ as $\v A = n_1  \v a_1 +n_2  \v a_2$ and $\v A' = n'_1  \v a_1 +n'_2  \v a_2$. Using the set of integers $n_1, n_2, n_1', n_2'$ and applying $\hat H$ to $K$ in the form (\ref{action}), we establish the more practical relation

\bea
\left[ \hat H -  i \frac{\partial}{\partial t}\right] K(n_1, n_2; n'_1, n'_2; t) = -i \delta(t) \delta_{n_1, n'_1} \delta_{n_2, n'_2}.
\label{II.11}
\eea
Now that the stage has been set, we may discuss particle statistics through canonical quantization.

\subsection{Discrete Feynman propagators in bosonic and fermionic quantization} \label{s2.2}

Our ideas can be applied easily to second quantization schemes. The purpose of this section is to derive the usual expressions for correlators and transition amplitudes in terms of the propagator previously defined. A discrete, non-relativistic version of a Feynman propagator shall be given as a commutator. We start with the usual assumptions of local field theory, but defined on lattices rather than continuous space. Promoting the localized functions $\xi_{\v A}, \xi_{\v A}^{*}$ to local field operators $F_{\v A}, F_{\v A}^{\dagger}$, we have a bilinear hamiltonian

\bea
H = \sum_{\v A, \v A'} H_{\v A, \v A' } F_{\v A }^{\dagger} F_{\v A'} 
\label{III.0}
\eea
where $H_{\v A, \v A'}$ are coupling coefficients satisfying $ H_{\v A, \v A' } =  H_{\v A', \v A }^{*}$. The hamiltonian (\ref{III.0}) contains as a particular case the periodic lattice with nearest-neighbor interactions:
\bea
H = \Delta \sum_{\v A, i}  F_{\v A + \bfalpha_i}^{\dagger} F_{\v A}  + \mbox{h.c.} + E_0 \sum_{\v A}  F_{\v A}^{\dagger} F_{\v A}. 
\label{III.1}
\eea
Bosonic or fermionic commutation relations can be imposed in the Schr\"odinger picture. We find it convenient to use the same symbol for both types of particles, but we distinguish the canonical commutation relations by subscripts:

\bea
\left[ F_{\v A}, F_{\v A'}^{\dagger} \right]_{\pm} = \delta_{\v A, \v A'}, \qquad \left[ F_{\v A}, F_{\v A'} \right]_{\pm} = 0,
\label{III.2}
\eea
where $(+)-$ stands for (anti) commutators. In the Heisenberg picture, the evolution equations are given by the commutator for both fermions and bosons

\bea
\frac{d F_{\v A}(t)}{dt} = i \left[ H , F_{\v A}(t) \right]. 
\label{III.2.1}
\eea
This in turn leads to solutions that are similar for both types of particles: With the aid of the two relations in (\ref{III.2}), it is possible to evaluate the r.h.s. of (\ref{III.2.1}) as a linear combination of field operators \footnote{There is a double sign flip for fermions in the commutators (\ref{III.2.1}), therefore (\ref{III.2.1.1.1}) is valid for fermions {\it and\ }bosons.}
\bea
 \left[ H , F_{\v A}(t) \right] = - \hat H  \left[ F_{\v A}(t) \right] = - \sum_{\v A'} H_{\v A, \v A'}   F_{\v A'}(t)
\label{III.2.1.1.1}
\eea
where we have applied $\hat H$ in the sense of (\ref{II.6.1}) and (\ref{action}). 

Upon integration over the variable $t$ in (\ref{III.2.1}), one has an evolution map provided by a single exponential operator, which is typical of bilinear hamiltonians:
\bea
F_{\v A}(t) = \left[ \sum_{n=0}^{\infty} \frac{ \left[-i(t-t') \right] ^n}{n!}  \hat H^n \right] \left[ F_{\v A}(t') \right].
\label{III.2.1.1}
\eea
As before, the application to $F_{\v A}$ on the r.h.s. of (\ref{III.2.1.1}) comes from the rule (\ref{action}). The operator in (\ref{III.2.1.1}) is nothing else than the matrix representation of $U$ and thus the propagator itself: 
\bea
 F_{\v A}(t) = \sum_{\v A'} K(\v A, \v A'; t-t') F_{\v A'}(t').
\label{III.2.2}
\eea
Finally, using (\ref{III.2.2}) and the (anti) commutation relations (\ref{III.2}) one finds that $K$ is related to the (anti) commutator at different times as

\bea
K(\v A, \v A'; t-t') = \left[ F_{\v A}(t), F_{\v A'}^{\dagger}(t') \right]_{\pm}.
\label{III.3}
\eea
This result is nothing else than the {\it discrete, non-relativistic Feynman propagator.\footnote{It is not necessary to take the expectation value between vacuum states in this derivation. The l.h.s. of (\ref{III.3}) is proportional to the identity operator in Fock space.}\ }Based on commutators of the type (\ref{III.3}), Lieb and Robinson \cite{lieb} proved general results for the speed of propagation in discrete space. Our task is to actually evaluate the l.h.s. of (\ref{III.3}). This shall be done in sections \ref{structural} and \ref{summary}.

\subsection{Evolution of many particles} \label{s2.3}

Now we turn to the evolution problem of multiparticle states. The transition probabilities are also related to the propagator in the following form. Denoting the in-site occupation numbers by $ \{ N_{\v A} \}_{\v A \in \rm{A}}$ and their corresponding states by $|\{ N_{\v A} \}\>$, we write the correlation amplitudes as the diagonal matrix elements of $U$, \ie

\bea
\< \{  N_{\v A}  \}, 0 | \{  N_{\v A}  \}, t \> = \< \{  N_{\v A}  \}| U_{t}|\{  N_{\v A}  \} \>,
\label{III.4}
\eea
while the transition probability is given by the more general matrix elements
\bea
\< \{  N'_{\v A}  \}| U_{t}|\{  N_{\v A}  \} \>.
\label{III.4.1}
\eea
For bosons, $ \{ N_{\v A} \}$ is a string of non-negative integers, while for fermions it is a string of $0$'s and $1$'s (we are not concerned now with their spin). We evaluate (\ref{III.4}) by means of the relation

\bea
 |\{ N_{\v A} \}\>= \prod_{\v A}\frac{ (F_{\v A})^{N_{\v A}}}{\sqrt{N_{\v A }!}} | \rm{vacuum}\>  
\label{III.5}
\eea
(valid at $t=0$) and the definition of the Heisenberg picture with a reversed time $F_{\v A}(-t) = U_{t} F_{\v A} U^{\dagger}_{t}$. Replacing these relations in (\ref{III.4.1}) and rearranging the field operators with (\ref{III.2}), we can show that the transition amplitudes can be cast in terms of products of propagators
\bea
\< \{  N'_{\v A}  \}, 0 | \{  N_{\v A}  \}, t \> =  \sum_{\{S_{\v A, \v A' }\}} \prod_{\v A, \v A'} \frac{\sqrt{ N_{\v A}! N'_{\v A}! } \left[K(\v A, \v A';t)\right]^{S_{\v A, \v A'}}}{S_{\v A, \v A'}!} 
\label{III.6}
\eea
for bosons, and the product over all occupied sites

\bea
\< \{  N'_{\v A}  \}, 0 | \{  N_{\v A}  \}, t \> =  \prod_{\v A, \v A'} K(\v A, \v A';t)
\label{III.7}
\eea
for fermions. Both (\ref{III.6}) and (\ref{III.7}) have total particle conservation as a selection rule. The usage of these formulae and their detailed derivation is given in appendix A. 

In this way, the computation of the propagators for different lattices opens the possibility of finding the diffusive behavior of many particles.

\section{Structural analysis and mathematical methods} \label{structural}

We would like to address the problem of finding explicitly the propagators previously defined. Their functional form is obviously determined by the specific type of lattice. However, there are certain generalities that we can prove and exploit for many types of such lattices. In the following we develop the necessary techniques for the evaluation of kernels, ranging from Clifford algebras in crystals to the application of special functions.

\subsection{Superalgebra and Dirac lattices} \label{s3}

In this section we establish a number of results on the algebraic properties of tight-binding hamiltonians, providing a connection with Clifford algebras. We focus on the implications of lattice symmetries which contain the operations of other types of lattices; for instance, the symmetry operators of hexagonal lattices are the {\it square root\ }of operators in a triangular crystal. This shall prove useful in the construction of kernels, in full analogy with the expression of the Dirac propagator in terms of the Klein-Gordon one \cite{bjorken}.

Hamiltonians of the type (\ref{I2}) allow a decomposition in terms of $su(2)$ operators and the members of another algebra, generally a non-compact one. Consider the following definitions of ladder or inter-lattice operators \footnote{The definitions (\ref{IV.0}), (\ref{IV.0.1}) and (\ref{IV.1}) contain $\bfalpha_1$ as the vector connecting sublattices A and B. This vector apparently lies in a privileged position, but it is possible to use any other $\bfalpha_i$ in compliance with the crystal symmetry.} 

\bea
\tau_{+} = \sum_{\v A} | \v A  \>\< \v A + \bfalpha_1 |, \quad \tau_{-} = \tau_{+}^{\dagger},
\label{IV.0}
\eea
and their vector form
\bea
\tau_1 = \tau_- + \tau_+, \quad \tau_2 = i(\tau_-  - \tau_+), \quad \tau_{3} =  \sum_{\v A} | \v A  \>\< \v A| -  \sum_{\v B} | \v B  \>\< \v B|.
\label{IV.0.1}
\eea
Consider also the definition of sublattice operations through
\bea
p_{-}= \sum_{\v A, i} | \v A  \>\< \v A +\bfalpha_i - \bfalpha_1 | + | \v A + \bfalpha_1 \>\< \v A +\bfalpha_i |, \quad p_{+} = p_{-}^{\dagger}.
\label{IV.1}
\eea
We have the following algebraic relations for the inter-lattice operators $\bftau$
\bea
\left[\tau_{+}, \tau_{-} \right] = \tau_3, \qquad \left[ \tau_{\pm}, \tau_{3} \right] = \mp 2 \tau_{\pm}, \qquad \{ \tau_i, \tau_j \} = 2 \delta_{ij} \v 1_2,
\label{IV.1.1}
\eea
while
\bea
\left[ p_{\pm}, \tau_{\pm} \right] = \left[ p_{\pm}, \tau_{3} \right] = 0,
\label{IV.1.2}
\eea
\bea
\left[ p_{+}, p_{-} \right] = 0,
\label{IV.2}
\eea
for the operators within sublattice A (alternatively B, depending on the convention). These algebraic properties can be checked easily by direct multiplication and a proper redefinition of summation indices, valid when the lattices are infinite. The meaning of $\tau_{\pm}, \tau_3$ and $p_{\pm}$ can be readily given from (\ref{IV.1.1}), (\ref{IV.1.2}) and (\ref{IV.2}): The set $\{ \tau_i/2 \}_{i=1,2,3}$ forms an $su(2)$ representation of spin $1/2$. It also forms a Clifford algebra. In the context of the Dirac equation \cite{bjorken}, this can be identified as the $su(2)^*$ summand in the algebra $su(2) \oplus su(2)^*$, which is isomorphic to the Lorentz $so(3,1)$. In the relativistic description, the spin associated to $su(2)^*$ (called $*$-spin) describes big and small components of spinors for spin up and down respectively. In crystals, the $*$-spin describes the probability amplitude of occupying lattice A or B. Moreover, the operators $p_{\pm}$ can be related to Bloch's quasi-momentum by applying them to Bloch states $|\phi(\v k) \>$ of sublattice A:

\bea
p_{\pm}|\phi(\v k) \> = p_{\pm} \sum_{\v A} e^{i \v k \cdot \v A}| \v A \> = \left[\sum_{i=1}^{c} e^{\pm i \v k \cdot (\bfalpha_i - \bfalpha_1)} \right] |\phi(\v k) \>.
\label{IV.3}
\eea
It should be noted that the relations (\ref{IV.1.1}) and (\ref{IV.1.2}) are quite general, but (\ref{IV.2}) and (\ref{IV.3}) depend strongly on the periodicity of the system\footnote{See e.g. \cite{sadurni} for a decomposition containing $su(2)$ and the Weyl algebra, which was proposed in the construction of Dirac-Moshinsky oscillators \cite{dmo} using non-periodic lattices.}. Some deformed lattices with non-homogeneous nearest-neighbor couplings can be obtained by the substitution $\Delta \mapsto \Delta(\v A)$, resulting in operators of the form 

\bea
\fl p_{-} \mapsto \sum_{\v A, i} \frac{\Delta(\v A)}{\Delta}\left( | \v A  \>\< \v A +\bfalpha_i - \bfalpha_1 | + | \v A + \bfalpha_1 \>\< \v A +\bfalpha_i |\right), \quad p_{+} = p_{-}^{\dagger}.
\label{IV.3.1}
\eea
This substitution may alter the commutator (\ref{IV.2}) by introducing non-trivial algebras of the type

\bea
\left[ p_{+}, p_{-} \right] = F(p_{+},p_{-})\neq 0,
\label{IV.4}
\eea
where $F$ is some hermitian function. Regardless of the explicit form of $F$, we can find the $S(2)$ superalgebra using the following definitions of hypercharges

\bea
Q_{+} = \tau_+ p_-, \quad Q_{-} = \tau_- p_+,
\label{sup1}
\eea
\bea
Q_{1} = Q_+ + Q_-, \quad Q_{2} = i( Q_+ - Q_- ) .
\label{sup2}
\eea
The hermitian supercharges (\ref{sup2}) satisfy

\bea
\left\{ Q_i , Q_j \right\} = 2 \delta_{ij} \v C, \qquad \left[ Q_i , \v C \right] = 0
\label{sup3}
\eea
where $\v C$ is the central charge. Such an operator can be related to the hamiltonian of a particle on the lattice in the form 
\bea
\v C = \left(H-\frac{E_1+E_2}{2}\right)^2 - \left(\frac{E_2-E_1}{2}\right)^2 
\label{central}
\eea
(see next section), which gives the resulting superalgebra the character of a dynamical supersymmetry. Although this was recognized in \cite{casta} in the context of the Dirac equation, the present treatment seems to be the first to recognize such a symmetry in the context of crystals.

Since we are interested in exact propagators for homogeneous lattices, we shall restrict ourselves to the case (\ref{IV.1}) in the following sections.

\subsection{Spectrum}
Now we cast the hamiltonian (\ref{I2}) in terms of the spectrum generating algebra (\ref{IV.1.1}), (\ref{IV.1.2}) and (\ref{IV.2}) in the form

\bea
H= \Delta \left[ \tau_{+} p_{-} + \tau_{-} p_{+} \right] + \mu \tau_3 + E_0 ,
\label{IV.5}
\eea
where the constants $\mu =( E_2- E_1)/2, E_0 =( E_2+ E_1)/2$ have been introduced. At this point we find it convenient to rescale the time variable in the Schr\"odinger equation $H |\quad\> = i \partial |\quad\> / \partial t$ by the substitution $t \mapsto t / \Delta $, $E_0 \mapsto \Delta E_0 $ and $\mu \mapsto \Delta \mu $, with the effect that $\Delta =1$ can be chosen without loss of generality. 

The stationary problem can be solved by computing the spectrum of the central charge $\v C = (H-E_0)^2 - \mu^2 $. For homogeneous lattices we have $F\equiv 0$, leading to

\bea
(H-E_0)^2 = p_+ p_- + \mu^2,
\label{IV.6}
\eea
which is a scalar, since it acts identically in both spinor components corresponding to the two sublattices. Denoting the eigenvalues of $p_+ p_-$ by $\epsilon_{\v k}^2$ and the eigenkets and eigenvalues of $H$ by $|\psi(\v k)\>$ and $E_{\v k}$ respectively, we find

\bea
E_{\v k} = E_0 \pm \sqrt{\epsilon_{\v k}^2 + \mu^2} \equiv E_0 \pm \lambda_{\v k}. 
\label{IV.7}
\eea
The two-component spinors $|\psi(\v k)\>$ are given in terms of the eigenkets $| \pm \>$ of $\tau_3$ by the rule 
\bea
|\psi(\v k)\> = A_{+}|+\>\otimes |\phi(\v k)\> + A_{-}|-\> \otimes |\phi(\v k)\> ,
\label{IV.8}
\eea 
with $A_{\pm}$ two complex coefficients that depend on $\v k$. We are now in the position to evaluate propagators by spectral decomposition.

\subsection{Spectral decomposition of the propagator} \label{spectral}

A very efficient method to obtain spinorial propagators comes from relativistic quantum mechanics, where the propagator of the Dirac equation can be related to the Klein-Gordon propagator by the application of the complex conjugate Dirac operator \cite{bjorken}. Similarly, our algebraic structure in a crystal has led us to an operator $(H-E_0)^2$ acting independently on each sublattice (a scalar problem) as opposed to $H-E_0$ acting on both sublattices and mixing them (a spinorial problem). Therefore we can solve the scalar problem and find the spinorial propagator as follows: Given a scalar kernel $K_S$ such that

\bea
\left[ \left(H- E_0 \right)^2 + \frac{\partial^2}{\partial t ^2} \right] K_S (\v A, \v A'; t) = -i \delta(t) \delta_{\v A, \v A'}
\label{V.1}
\eea
we can obtain

\bea
e^{iE_0 t}\left[ H-  i\frac{\partial}{\partial t} \right]e^{-iE_0 t}\left[ H +  i\frac{\partial}{\partial t} - E_0  \right] K_S (\v A, \v A'; t) = -i \delta(t) \delta_{\v A, \v A'},
\label{V.2}
\eea
which, multiplied by $e^{-iE_0 t}$ on both sides, is equivalent to

\bea
\left[ H-  i\frac{\partial}{\partial t} \right] \left\{e^{-iE_0 t} \left[ H +  i\frac{\partial}{\partial t} - E_0  \right] K_S (\v A, \v A'; t) \right\} = -i \delta(t) \delta_{\v A, \v A'},
\label{V.3}
\eea
and this leads to a spinor propagator $K$ given by the expression between braces

\bea
 K(\v A, \v A'; t)=e^{-iE_0 t} \left[ H +  i\frac{\partial}{\partial t} - E_0  \right] K_S (\v A, \v A'; t).
\label{V.4}
\eea
The scalar propagator $K_S$ can be obtained via spectral decomposition, represented by an integral over a suitable contour $C$:

\bea
K_S (\v A, \v A'; t) &=& - \frac{i}{2 \pi} \int_{C} dE \int_{\Omega_B} d \v k \frac{\phi_{\v k}(\v A) \phi^*_{\v k}(\v A') e^{-i E t}}{\lambda^2_{\v k}- E^2} \nonumber \\
&=& \int_{\Omega_B}d \v k \frac{\phi_{\v k}(\v A) \phi^*_{\v k}(\v A')}{2 \lambda_{\v k}} \left[ e^{-i \lambda_{\v k} t} - e^{i \lambda_{\v k} t} \right].
\label{V.5}
\eea
\begin{figure}[!h] \begin{center}  \includegraphics[scale=0.5]{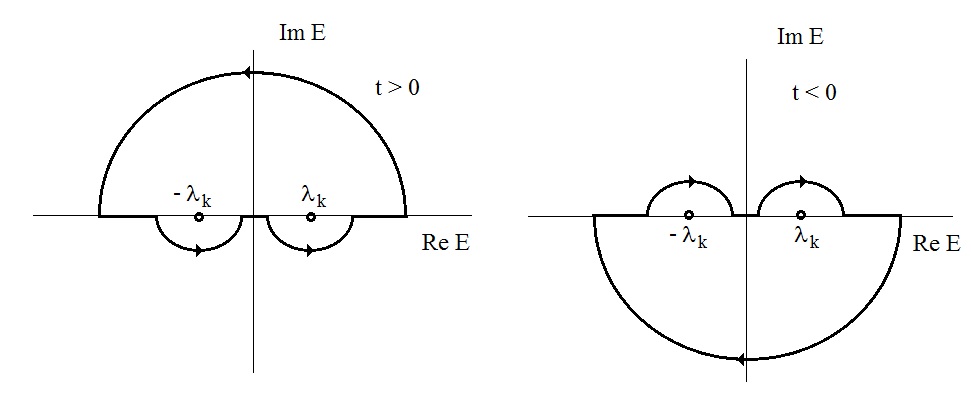} \end{center} 

\caption{ \label{fig:c}Possible contours of integration in the calculation of $K_S$ in (\ref{V.5}). Both paths enclose the negative and positive poles, as they are necessary for the spectral decomposition of the propagator. The path on the left is our definite choice, giving forward propagation. The one on the right is valid only for $t<0$ and does not contribute to the kernel. Wick's angle should not be introduced in this kind of application.} \end{figure}

Here we have made an explicit separation of the exponentials between brackets, as their presence depends on the contour of integration $C$, depicted in figure \ref{fig:c}. The choice of $C$ in solid state physics and photonic crystals is quite different from the one used in relativistic quantum mechanics: in the present context we are sure that all the energy eigenvalues of our problem are {\it physical\ }and should contribute to the spectral representation of our {\it unitary\ }kernel. For this reason, $C$ encloses both poles $\pm \lambda_{\v k}$ as opposed to relativistic contours (e.g. Klein-Gordon) where by hand one introduces Wick's angle in order to make an explicit separation of contributions from particles and antiparticles. Our kernels are non-relativistic objects and must contain a complete set of solutions for $t>0$. 

With these considerations in mind, we refer the reader to our tables of results \ref{tab:1}, \ref{tab:2} for propagators in hexagonal and dimeric lattices, which can be constructed in terms of triangular and monomeric propagators respectively. In the following sections we focus on the evaluation of (\ref{V.5}).

\subsection{Mass and generating functions in operational form}

We can compute the integral in (\ref{V.5}) using a trick based on generating functions. Such functions have the spectral gap $\mu$ as expansion parameter. In the relativistic language, the parameter $\mu$ represents a rest mass. Consider the integral

\bea
K_S (\v A, \v A'; t)= \int_{\Omega_B}d \v k \phi_{\v k}(\v A) \phi^*_{\v k}(\v A')  \frac{\sin \left( \lambda_{\v k} t \right)}{ i \lambda_{\v k}} .
\label{VI.1}
\eea
Each factor in the integrand can be put in a convenient form. The factor depending on time and energy can be written in terms of a translation operator of the variable $\mu^2$ as
\bea
\frac{\sin \left( \lambda_{\v k} t \right)}{ i \lambda_{\v k}}=\exp \left[ -i \epsilon_{\v k}^2 \left( \frac{i}{2\mu} \frac{\partial}{\partial \mu} \right)  \right] \frac{\sin \left( \mu t \right)}{i \mu}. 
\label{VI.2}
\eea
For Bloch waves  $\phi(\v A)$ we simply substitute $e^{i \v k \cdot \v A} / \sqrt{\omega}$ in (\ref{VI.1}), while the energy $\epsilon_{\v k}^2$ corresponds to the eigenvalue of the tight-binding hamiltonian of the sublattice A and it is given typically by a trigonometric function of $\v k$. Replacing (\ref{VI.2}) and the Bloch waves back in (\ref{VI.1}) we have

\bea
\fl K_S (\v A, \v A'; t)= \frac{1}{\omega} \left\{\int_{\Omega_B}d \v k \exp \left[i \v k \cdot (\v A - \v A') - i \epsilon_{\v k}^2 \left( \frac{i}{2\mu} \frac{\partial}{\partial \mu} \right)  \right] \right\} \frac{\sin \left( \mu t \right)}{i \mu}.
\label{VI.3}
\eea
Occasionally, the function $\epsilon^2_{\v k}$ will contain constant terms that may be subtracted for convenience. We denote such a constant by $\gamma$ and obtain (\ref{VI.3}) in the more general form
\bea
\fl K_S (\v A, \v A'; t)= \frac{1}{\omega} \left\{\int_{\Omega_B}d \v k \exp \left[i \v k \cdot (\v A - \v A') - i (\epsilon_{\v k}^2 - \gamma) \left( \frac{i}{2\mu} \frac{\partial}{\partial \mu} \right)  \right] \right\} \frac{\sin \left( t \sqrt{\mu^2+\gamma}  \right)}{ i \sqrt{\mu^2+\gamma}}. \nonumber \\
\label{VI.3.1}
\eea
Finally, $\epsilon_{\v k}$ and the operator between brackets are determined by specific lattice realizations:
\subsection{Hexagonal kernel}
As an example, we can show that hexagonal lattices with a spectral gap possess propagators which can be put in terms of triangular kernels. Let us denote the interlattice vectors $\bfalpha_i$ by $\v b_1, \v b_2, \v b_3$ (see section \ref{hexagon}) and the triangular primitive vectors by $\v a_1, \v a_2$. With the well-known dispersion relation (\ref{IV.7}), we obtain explicitly \cite{semenoff, wallace}

\bea
\fl \epsilon_{\v k}^2 = \left| \sum_{i=1}^{3}\exp (i \v k \cdot \v b_i)\right|^2 = 3 + 2 \left[\cos (\v k \cdot \v a_1) + \cos (\v k \cdot \v a_2) + \cos (\v k \cdot \v a_1 + \v k \cdot \v a_2)\right],
\label{VI.4}
\eea
which is similar to the exponent in the generating function of the two-index Bessel $J_{m,n} (x;-i)$. Such functions were studied by Dattoli \cite{1.3}. The generating function of variable $x$ has the form

\bea
\fl \exp \left\{ i x \left[ \cos k_1 + \cos k_2 + \cos (k_1 + k_2)\right] \right\} = \sum_{m,n \in \v Z} e^{im (k_1+ \pi/2)} e^{in (k_2+ \pi/2)} J_{m,n} (x;-i).
\label{VI.4.1}
\eea
Let us take $\gamma=3$; integrating over $\Omega_B$ in (\ref{VI.3}) and using the completeness relation of Bloch waves leads to a single Bessel function of two indices, whose argument $x =\frac{i}{\mu} \frac{\partial}{\partial \mu}$ acts on a function of the parameter $\mu$. In order to show this, we perform the change of variables $k_i = \v k \cdot \v a_i$ in reciprocal space, mapping each sector of $\Omega_B$ into a square of area $4 \pi^2$. This transforms integrals of the type (\ref{VI.3.1}) in the following way:

\bea
\fl \frac{1}{\omega} \int_{\Omega_B}d \v k \exp \left\{i \v k \cdot \v A  - i \left[ \cos (\v k \cdot \v a_1) + \cos (\v k \cdot \v a_2) + \cos (\v k \cdot \v a_1 + \v k \cdot \v a_2)\right] \left[ \frac{i}{\mu} \frac{\partial}{\partial \mu} \right] \right\}  \nonumber \\  
\fl = \frac{1}{(2 \pi)^2} \int_{0}^{2\pi} \int_{0}^{2\pi} dk_1 dk_2 \quad e^{i (n_1 k_1+ n_2 k_2)} e^{ - i  \left[ \cos k_1 + \cos k_2 + \cos (k_1 + k_2)\right] \left[ \frac{i}{\mu} \frac{\partial}{\partial \mu} \right] } \nonumber \\  \fl = i^{-n_1 - n_2} J_{n_1,n_2} \left( \frac{i}{\mu} \frac{\partial}{\partial \mu} ; -i\right),
\label{VI.4.2}
\eea
which is indeed a compact result. We refer the reader to formulae (\ref{5.3}) and (\ref{5.4}) for the final expression of the hexagonal propagator. We may further show that the triangular kernel without spectral gap and at time $t$ can be given directly by the aforementioned Bessel functions $J_{m,n}(2t)$, see formula (\ref{4.2}).
\subsection{Dimeric kernel}
Yet a simpler example comes in the form of dimeric chains. We replace $\v k \mapsto k$, $\Omega_B \mapsto [0,2\pi]$, $\gamma \mapsto 2$ and the dispersion relation
\bea
\epsilon_{k}^2 = 4 \cos^2  k = 2 [1 + \cos (2 k ) ]
\label{VI.5}
\eea
in (\ref{VI.3.1}). The resulting one-dimensional version of the operator in (\ref{VI.2}) is

\bea
\exp \left\{ i  k  (n-n') -  i \cos (2 k ) \left[ \frac{i}{\mu} \frac{\partial}{\partial \mu} \right]  \right\}
\label{VI.6}
\eea
and it contains the Bessel generating function of the first kind \footnote{This is the Jacobi-Anger expansion \cite{1}. Bessel functions (\ref{VI.8}) evaluated at operators are understood in terms of their ascending series. }:

\bea
e^{i x \cos(2k)} = \sum_{n \in \v Z} i^n e^{2ink} J_n(x).
\label{VI.7}
\eea
Therefore, the integration over the first Brillouin zone and the completeness relation of Bloch waves leads to a single Bessel function with argument $ x = \frac{i}{\mu} \frac{\partial}{\partial \mu}$ and indices $n-n'$. In order to see this, we evaluate the following integral: 

\bea
\fl \frac{1}{2 \pi} \int_{0}^{2 \pi} dk e^{ink} \exp \left\{  -  i  \cos (2 k ) \left[ \frac{i}{\mu} \frac{\partial}{\partial \mu} \right]   \right\}
= i^{-n} \cases{J_{n} \left( \frac{i}{\mu} \frac{\partial}{\partial \mu} \right) & for $n$ even \\ 0 & otherwise }
\label{VI.8}
\eea
For the final result of the propagator, see formula (\ref{2.3}), which gives the dimeric kernel in terms of a homogeneous chain kernel. Furthermore, the latter possesses a propagator given directly by the Bessel function $J_n(t)$ as in formula (\ref{1.3}).

It is important to mention that our operational expressions are exact and provide useful expansions whenever $\mu$ is a large parameter. The propagators given in this manner can be turned easily into simpler functions by using expansions truncated at arbitrary order $-$for example, the ascending form of the Bessel operator (\ref{VI.8})$-$ while the direct use of spectral decompositions makes this task computationally prohibitive. It is in this sense that the ascending series of Bessel operators of one or two indices become useful. 

\subsection{Strong gap expansions of Bessel operators}
In certain limits there is yet a more suitable way to express (\ref{VI.4.2}) or (\ref{VI.6}). We invoke a very general identity for operators: Let $f$ and $g$ be analytic functions. We have proven in the appendix B of \cite{5} the following formula

\bea
 f\left( \frac{d}{d \lambda} \right) g(\lambda) =g(\lambda) f \left( \frac{g'(\lambda)}{g(\lambda)} \right) + \frac{g(\lambda)}{2} \left(\frac{g'(\lambda)}{g(\lambda)} \right)'  f'' \left( \frac{g'(\lambda)}{g(\lambda)} \right) + \dots
\label{b5}
\eea
This formula in turn can be applied to our problem with $\lambda=\mu^2$ and the appropriate substitutions for $f$ and $g$. It allows truncability when $\mu \gg 1$. In fact, we have used this expression \cite{5} in the problem of diffraction by edges inside a periodic medium. There we could relate the first term in (\ref{b5}) with a paraxial wave approximation. Here, on the other hand, we have a leading term that is related to large gaps in the spectrum. We give the propagators resulting from this expansion in (\ref{2.5}) and (\ref{2.7}) for chains and (\ref{5.3}) applied to (\ref{5.6}) for hexagonal lattices. 

 In addition to the operational treatment given above, there are other expressions of our kernels that come in handy for different situations. We discuss this in the next section.

\subsection{The unexpected appearence of spherical waves}

The kernel of the dimer chain has an alternate expansion arising from exponentials of radicals. Surprisingly, expansions in spherical Bessel functions appear in this one-dimensional problem. The usefulness of such expansions resides in their fast convergence, as well as in the form of the expansion parameters (not a power series in $\mu$). Consider the spherical wave expansion \cite{abramowitz} of vectors $\v r, \v r'$ and $\cos \theta = \hat \v r \cdot \hat \v r'$:

\bea
\fl \frac{\exp \left\{ i \kappa \sqrt{r^2+r'^2-2 r r' \cos \theta} \right\} }{i \kappa \sqrt{r^2+r'^2-2r r' \cos \theta}} &=& 4 \pi \sum_{l=0}^{\infty} j_{l}(\kappa r_{>}) h^{(1)}_{l}( \kappa r_{<}) \sum_{m=-l}^{l} Y^{*}_{lm}(\hat \v r')Y_{lm}(\hat \v r) \nonumber \\ &=& \sum_{l=0}^{\infty} (2l+1) j_{l}(\kappa r_{>}) h^{(1)}_{l}(\kappa r_{<}) P_{l}(\cos \theta).
\label{VII.1}
\eea
This identity can be compared with the exponential functions of $\epsilon^2_{\v k}$ appearing in the spectral decomposition of scalar propagators (\ref{V.5}). For instance, the one dimensional version is given by

\bea
 K_S(n,m;t) &=& \frac{1}{2 \pi}\int_{0}^{2\pi} dk \quad e^{i(n-m)k} \left[\frac{\sin (t\sqrt{\mu^2+4\cos^2 k})}{ i \sqrt{\mu^2+4\cos^2 k}}\right].
\label{VII.1.2}
\eea
Evidently, the integrand contains exponentials of the form
\bea
\fl \frac{\exp \left\{ it \sqrt{\mu^2+ 4 \cos^2 k} \right\} }{\sqrt{\mu^2+ 4 \cos^2 k}} &=& \frac{\exp \left\{ it \sqrt{\mu^2+ 2+2 \cos(2 k)} \right\} }{\sqrt{\mu^2+ 2+2 \cos(2 k)}} \nonumber \\ &=& 4 \pi i t\sum_{l=0}^{\infty} j_{l}(\mu_{-}t) h^{(1)}_{l}(\mu_{+}t) \sum_{m=-l}^{l} (-1)^m e^{-2imk} \left[ P_l^{m}(0) \right]^2 
\label{VII.2}
\eea
where $P_l^{m}(0)$ is the normalized\footnote{We simplify the notation by choosing the normalization of $P_{l}^{m}$ such that $Y_l^m(\hat \v r_i) = P_l^m(\cos \theta_i) e^{im\phi_i}$.} associated Legendre polynomial evaluated at the origin and $\mu_{\pm} \equiv (\sqrt{\mu^2+4} \pm \mu)/2$. In this formula we have identified $t$ with $\kappa$, $2k+\pi$ with $\theta$ and $r_{>,<}$ with $\mu_{+,-} $. Furthermore, we have set $\hat \v r = \v i$ and $\hat \v r' = - \v i \cos \theta + \v j \sin \theta $. Using (\ref{VII.2}) in the integration of (\ref{VII.1.2}), one ends with an expansion of $K_S$ in terms of spherical waves, \ie 

\bea
\fl K_S(n,m;t) = \left(\frac{1+(-1)^{n+m}}{2} \right) (-1)^{\frac{|n-m|}{2}+1} \quad 4 \pi i t \sum_{l=|n-m|/2}^{\infty} j_l(\mu_+ t) j_l(\mu_- t) \left[ P_l^{\frac{|n-m|}{2}}(0) \right]^2. \nonumber \\ 
\label{VII.3}
\eea
The results obtained in this fashion are of fast convergence, due to factorials in the denominator of the ascending series for each spherical Bessel function. Inverses of factorials also appear in the associated Legendre function evaluated at the origin. Therefore, a few terms should be enough to obtain a good approximation of $K_S$. It is also important to note that (\ref{VII.3}) is not a power expansion of $\mu$, although it may serve well this purpose by expanding $\mu_{\pm}$ and re-expanding $j_l$. See formulae (\ref{2.4}) and (\ref{2.6}) for an application of this type in the propagator of the dimeric chain.

\subsection{Graphene, the gapless limit and Bessel polynomials} \label{graphene}
Our methods so far have led us to convenient expansions of the propagators, either scalar or (quasi) spinorial. However, the limit $\mu \rightarrow 0$ in the hexagonal case deserves a particular treatment. Its relevance comes from a comparison of the type graphene vs boron nitride: We would like to write down an explicit propagator for hexagonal lattices without spectral gap.

A close look into the corresponding dispersion relation suggests that the {\it spherical\ } analogues of the two-index Bessel functions should provide a framework akin to (\ref{VII.3}). However, such mathematical functions have not been defined in the literature and we shall not proceed in this direction. Instead, we tackle the problem in a more direct manner:

The ascending series of the two-index Bessel operator (\ref{VI.4.2}) has the form

\bea
J_{n,m} \left(\frac{i}{\mu} \frac{\partial}{\partial \mu}; -i \right) = \sum_{s=0}^{\infty} C_s^{nm} \left( \frac{i}{\mu} \frac{\partial}{\partial \mu} \right)^s
\label{VIII.1}
\eea
where the coefficients $C_s^{nm}$ are computed explicitly in appendix B. The function on which this operator acts is given in (\ref{VII.3}) as $\sin \left( t \sqrt{\mu^2 + \gamma} \right)/i\sqrt{\mu^2 + \gamma}$, which can be written in the form

\bea
\frac{i}{t \mu}\frac{\partial}{\partial \mu} \cos \left( t \sqrt{\mu^2 + \gamma} \right),
\label{VIII.2}
\eea
 eliminating thus the radical in the denominator. Therefore, it is sufficient to analyze the ascending series of the more general case

\bea
J_{n,m} \left(\beta \frac{\partial}{\partial \lambda}; -i \right) \exp \left( it\sqrt{\lambda + \gamma} \right)
\label{VIII.3}
\eea
with $\beta, \gamma$ two constants and $\lambda = \mu^2$ as before. Note that the gapless limit corresponds now to $\lambda \rightarrow 0$. The key point in this calculation comes from the generating function of the Bessel polynomials \cite{polynomials}, which is similar to the exponential in (\ref{VIII.3}):

\bea
\exp \left( it\sqrt{\lambda + \gamma} \right) = \sqrt{\frac{2t\sqrt{\gamma}}{i\pi}} \sum_{q=0}^{\infty} \frac{1}{q!} \left(\frac{it\lambda}{2\sqrt{\gamma}} \right)^q  K_{q-1/2}(-it\sqrt{\gamma})
\label{VIII.4}
\eea
where $K$ is a modified Bessel function of the second kind. Combining (\ref{VIII.1}), (\ref{VIII.3}) and (\ref{VIII.4}) we have

\bea
\fl J_{n,m} \left(\beta \frac{\partial}{\partial \lambda}; -i \right) \exp \left( it\sqrt{\lambda + \gamma} \right)= \nonumber \\ \sqrt{\frac{2t\sqrt{\gamma}}{i\pi}} \sum_{s=0}^{\infty} \sum_{q=0}^{\infty} \frac{(q)_s C_{s}^{nm}}{q!} \left(\frac{it}{2\sqrt{\gamma}} \right)^q  K_{q-1/2}(-it\sqrt{\gamma})    \beta^s  \lambda^{q-s}
\label{VIII.5}
\eea
with $(q)_s$ the Pochhammer symbol. Taking $\lambda=0$ leaves us with the remarkable expansion

\bea
\fl \left. J_{n,m} \left(\beta \frac{\partial}{\partial \lambda}; -i \right) \exp \left( it\sqrt{\lambda + \gamma} \right) \right|_{\lambda=0}= \sqrt{\frac{2t\sqrt{\gamma}}{i\pi}}  \sum_{q=0}^{\infty}  C_{q}^{nm} \left(\frac{it \beta}{2\sqrt{\gamma}} \right)^q  K_{q-1/2}(-it\sqrt{\gamma}).  
\label{VIII.6}
\eea
Finally, using (\ref{VI.3.1}), (\ref{VI.4.2}) and the appropriate substitutions $\beta=2i$, $\gamma=3$, we can relate the scalar propagator for graphene with the expression (\ref{VIII.3}) and the expansions (\ref{VIII.5}), (\ref{VIII.6}). With identities of the type $K_{\nu}(-iz)=\frac{2i^{-\nu-1}}{\pi}H^{(1)}_{\nu}(z)$ and trivial manipulations of the general term of the series, we obtain

\bea
\fl K_S(\v A, \v A'; t) &=& i^{n'_1 + n'_2 -n_1 - n_2} \left\{ \left. J_{n_1-n'_1,n_2-n'_2} \left(\beta \frac{\partial}{\partial \lambda}; -i \right) \frac{\sin \left( t \sqrt{\lambda + \gamma} \right)}{i\sqrt{\lambda + \gamma}}   \right\} \right|_{\lambda=0} \nonumber \\
&=& i^{n'_1 + n'_2 -n_1 - n_2} \left\{ \left. J_{n_1-n'_1,n_2-n'_2} \left(\beta \frac{\partial}{\partial \lambda}; -i \right) \frac{2i}{t }\frac{\partial}{\partial \lambda} \cos \left( t \sqrt{\lambda + \gamma} \right)  \right\} \right|_{\lambda=0} \nonumber \\
\fl &=&  i^{n'_1 + n'_2 -n_1 - n_2}  \sqrt{\frac{8t}{\pi^3 \sqrt{\gamma}}}  \sum_{q=0}^{\infty}  C_{q}^{n_1-n_1',n_2-n_2'} \left(\frac{\beta t }{2\sqrt{\gamma}} \right)^{q}  J_{q+1/2}(\sqrt{\gamma}t) \nonumber \\
\fl &=&  i^{n'_1 + n'_2 -n_1 - n_2}  \sqrt{\frac{8t}{\pi^3 \sqrt{3}}}  \sum_{q=0}^{\infty}  C_{q}^{n_1-n_1',n_2-n_2'} \left(\frac{i t }{\sqrt{3}} \right)^{q}  J_{q+1/2}(\sqrt{3}t).
\label{VIII.7}
\eea
The full spinorial kernel is obtained by means of (\ref{V.4}) and we include it in our summary of results, see (\ref{5.3}) and (\ref{5.4}).

\begin{figure}[!h] \begin{center} \begin{tabular}{ccc} \includegraphics[scale=0.25]{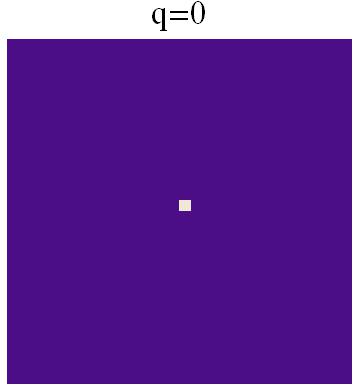} &  \includegraphics[scale=0.25]{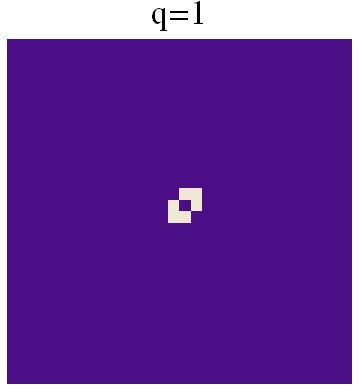} &  \includegraphics[scale=0.25]{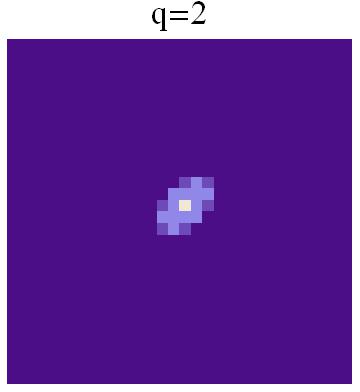} \\   \includegraphics[scale=0.25]{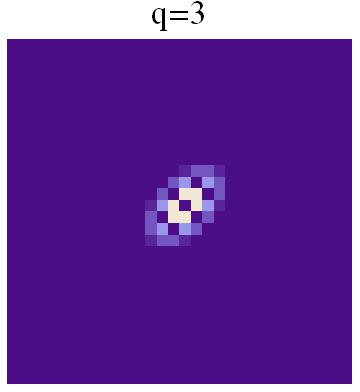} &  \includegraphics[scale=0.25]{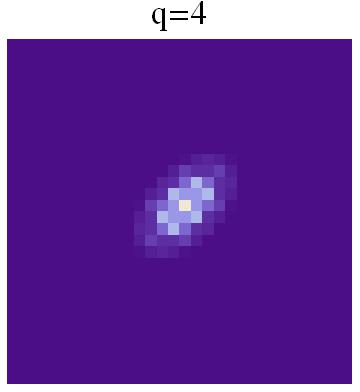} &  \includegraphics[scale=0.25]{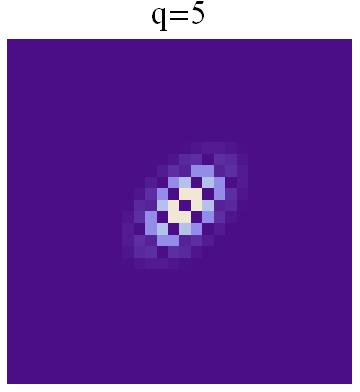} \\  \includegraphics[scale=0.25]{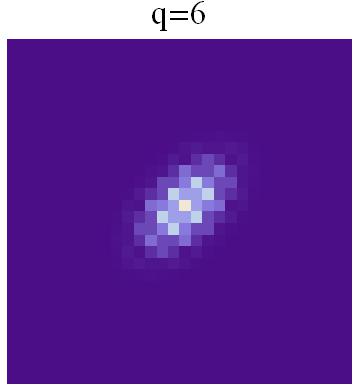} &  \includegraphics[scale=0.25]{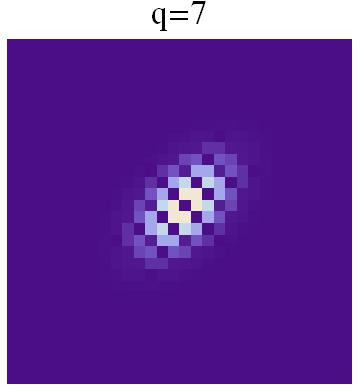} &  \includegraphics[scale=0.25]{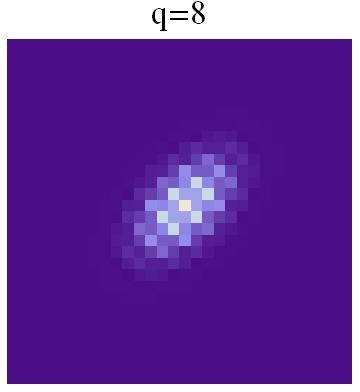}  \end{tabular} \end{center} 

\caption{ \label{fig:1} Density plot of the coefficients $C_q^{nm}$ for each value of $q$, where $n$ is the abscissa and $m$ is the ordinate. The dark areas indicate that the coefficient vanishes for values of $m,n$ beyond $q$. This in turn shows that the expansion (\ref{VIII.7}) can be truncated to order $q$ if $t \sim q$. } \end{figure}

We should note that (\ref{VIII.7}) is a quickly converging expansion due to an exponentially decreasing $C_q^{nm}$ with $q$ (see the factor $2^{-s}$ in (\ref{b.6})). From the physical point of view, we can further discuss the behavior of $|K_S|^2$ in space by recognizing that the dependence on the lattice integers falls completely on the coefficients $C_q^{nm}$. The values of $n$ and $m$ are limited by $q$, as shown by formula (\ref{b.6}) with the restrictions (\ref{b.4}). Also, each term of (\ref{VIII.7}) indicates that the major contribution is dictated by the caustic of the Bessel function $J_{q+1/2}$, resulting in $q+ 1/2 \sim t$. This argumentation finally shows that the occupied lattice points with integers $n,m$ (equivalently $n_1-n'_1,n_2-n'_2$) are limited by $t$ itself: beyond such values the probability amplitude vanishes exponentially, in accordance with a {\it classically forbidden region.\ }See figure \ref{fig:1}. 

Now that we have shown the techniques for the evaluation of kernels, we may discuss some applications.

\section{Application to diffusion} \label{s4}

The spreading of real wavepackets as a function of time has natural consequences on diffusion. Our interest is to present illustrative examples of diffusive processes taking place in crystals. The effects of the periodic structure are already visible in apparently naive situations where a single site is occupied at time $t=0$; in this kind of scenario our propagators become most useful. But more general configurations can be described as well: it is possible to place several particles at various sites of the crystal and study their evolution by means of the transition amplitudes (including correlations) computed in appendix A. Particle statistics play a fundamental role in the outcome of such processes. The results of the calculations can be compared for different types of lattices (one- and two-dimensional), showing that the effects in question depend on the coordination number. 

\subsection{Point-like stimuli}

As the simplest application, we consider point-like distributions at $t=0$ placed at an arbitrary site of the lattice. Then we study the distribution at later times. The probability distributions are simply given by $|K_{\v A, 0}(t)|^2$ for each $K$ and the expansion of these `pulses' or `pixels' can be compared with the famous Lieb-Robinson bound for propagation velocities \cite{lieb}. This gives rise to an increase in the probability density as a function of space-time in the form of a wavefront. Although this type of result can be derived with the method of caustics for integrals in the reciprocal lattice, here we present the densities resulting from full calculations of kernels. This makes visible the interference patterns occurring in the regions enclosed by the expanding wavefronts.

\begin{figure}[!h] \begin{center} \begin{tabular}{ccc} \includegraphics[scale=0.28]{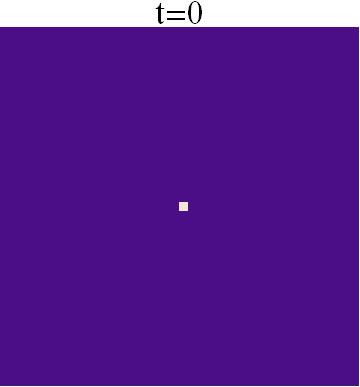} &  \includegraphics[scale=0.28]{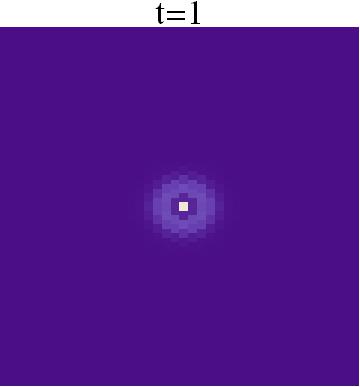} &  \includegraphics[scale=0.28]{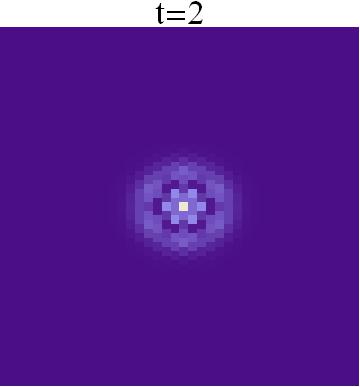} \\   \includegraphics[scale=0.28]{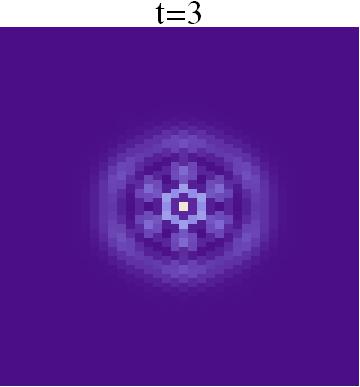} &  \includegraphics[scale=0.28]{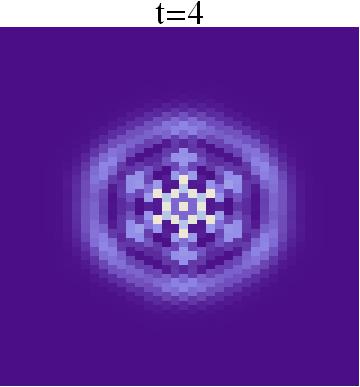} &  \includegraphics[scale=0.28]{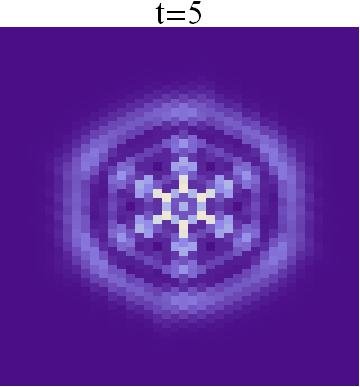} \end{tabular} \end{center} 

\caption{ \label{fig:2} Density plot of the probability distribution, emerging from a point-like initial condition in a triangular lattice with a $40\times40$ grid. Each panel shows the progressive expansion of the distribution. This is a visualization of the square modulus of the triangular kernel itself, given by the two-index Bessel function. The coordination number of the lattice is 6, which gives the pattern a peculiar ressemblance to a snow flake.} \end{figure}

\begin{figure}[!h] \begin{center} \begin{tabular}{ccc} \includegraphics[scale=0.28]{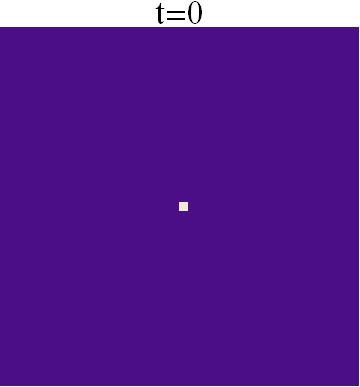} &  \includegraphics[scale=0.28]{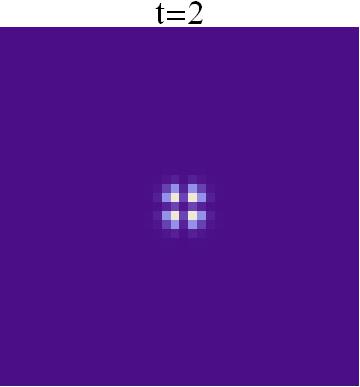} &  \includegraphics[scale=0.28]{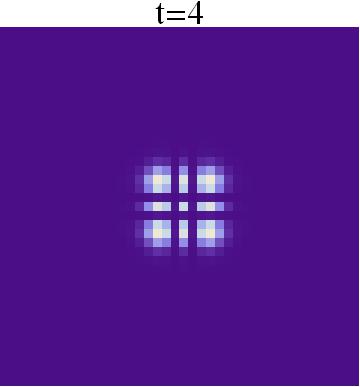} \\   \includegraphics[scale=0.28]{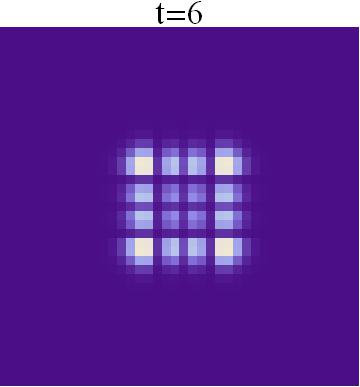} &  \includegraphics[scale=0.28]{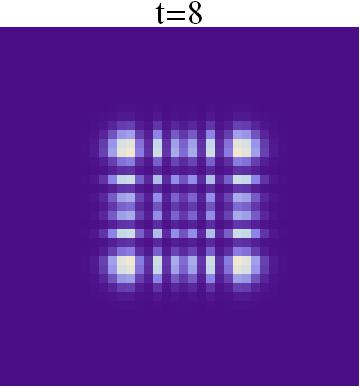} &  \includegraphics[scale=0.28]{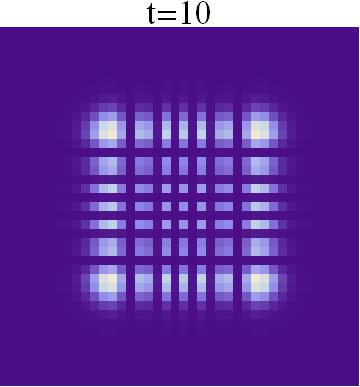} \end{tabular} \end{center} 

\caption{ \label{fig:3} Density plot of the probability distribution, emerging from a point-like initial condition in a square lattice and using a $40\times40$ grid. Each panel shows the progressive expansion of the distribution. This is also a visualization of the product of two Bessel functions. We note that the expansion occurs faster than in the hexagonal case. This is to be expected, since the coordination number in this case is 4.} \end{figure}

\begin{figure}[!h] \begin{center} \begin{tabular}{ccc} \includegraphics[scale=0.37]{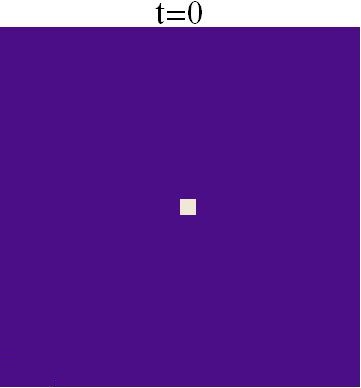} &  \includegraphics[scale=0.37]{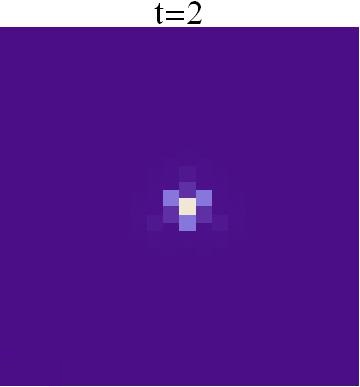} &  \includegraphics[scale=0.37]{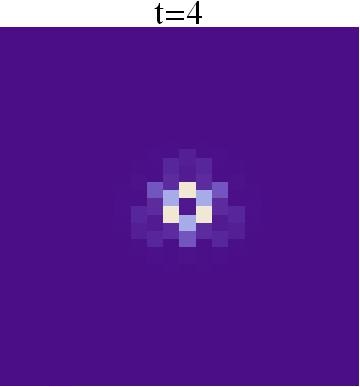} \\   \includegraphics[scale=0.37]{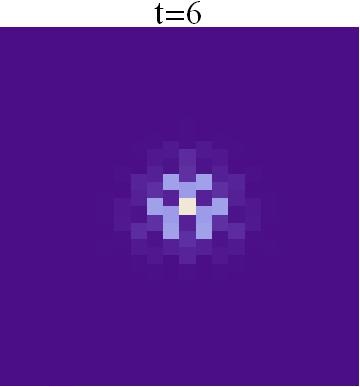} &  \includegraphics[scale=0.37]{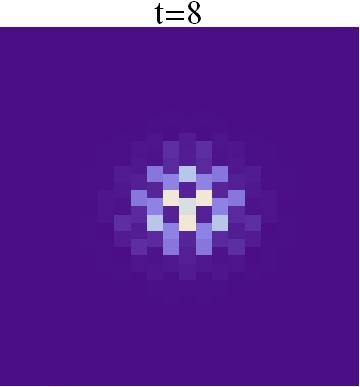} &  \includegraphics[scale=0.37]{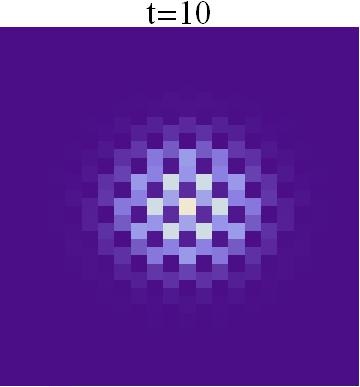} \end{tabular} \end{center} 

\caption{ \label{fig:4} Density plot of the probability distribution, emerging from a point-like initial condition in a homogeneous hexagonal lattice and using a $15\times15$ grid. Each panel shows the progressive expansion of the distribution. This is a visualization of the modulus of a gapless hexagonal propagator. Since the coordination number is 3 (minimal for 2d arrays), the expansion for this case is the slowest possible.} \end{figure}

\begin{figure}[!h] \begin{center} \begin{tabular}{ccc} \includegraphics[scale=0.28]{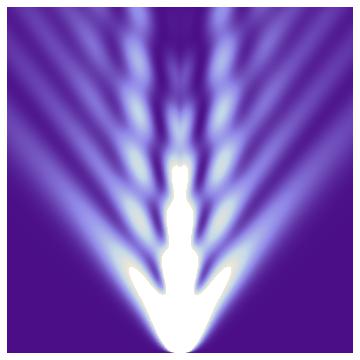} &  \includegraphics[scale=0.28]{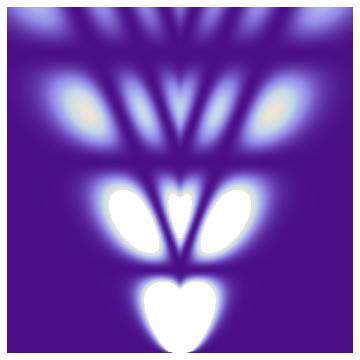} &  \includegraphics[scale=0.28]{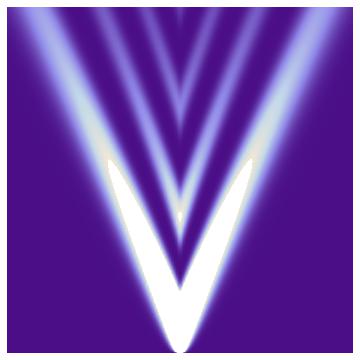} \end{tabular} \end{center} 

\caption{ \label{fig:5} Comparison of probability densities in space (abscissa) and time (ordinate). The left panel shows the evolution of the point-like initial condition using the triangular kernel along the primitive vector $\v a_1$, the central panel shows the corresponding result for a square lattice along the primitive vector $\v i$ and the right panel shows the evolution in a linear chain. Although the three problems exhibit fold-type caustics and a maximal speed of propagation, the structure of the interference region between caustics varies for each case.} \end{figure}

The results in sheets are shown in figures \ref{fig:2}, \ref{fig:3} and \ref{fig:4}. All the cases correspond to homogeneous lattices with only one atomic species, \ie no spectral gap. The shape of the wavefronts clearly changes as a function of the lattice coordination. Collapse and revivals also occur in assorted patterns. Another view of these effects can be found in figure \ref{fig:5}, where the evolution along a primitive vector is shown in space and time.

\begin{figure}[!h] \begin{center} \begin{tabular}{ccc} \includegraphics[scale=0.28]{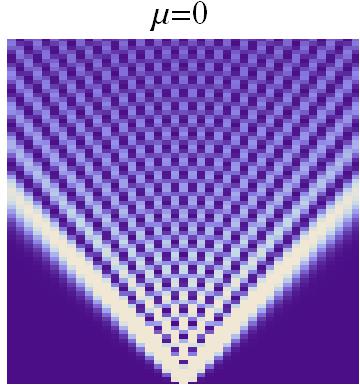} &  \includegraphics[scale=0.28]{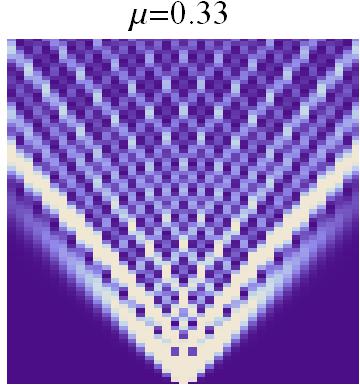} &  \includegraphics[scale=0.28]{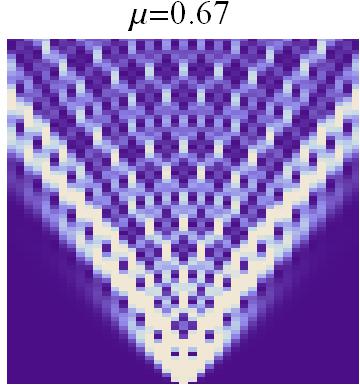} \\   \includegraphics[scale=0.28]{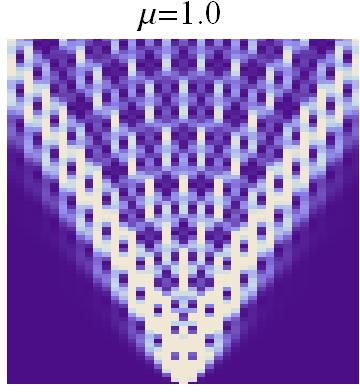} &  \includegraphics[scale=0.28]{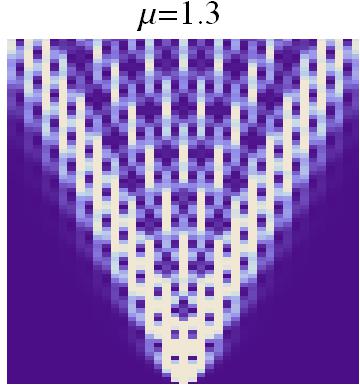} &  \includegraphics[scale=0.28]{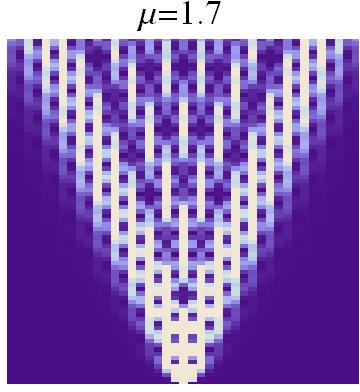} \end{tabular} \end{center} 

\caption{ \label{fig:6} The effects of an increasing spectral gap. The panels show the probability distributions emerging from a point-like initial condition as a function space (abscissa) and time (ordinate). The spectral gap $\mu$ is on top of each panel. As the value of $\mu$ increases, the slopes of the caustics also increase, indicating that the expansion of the point-like pulse becomes slower. The details of the interference region between the caustics are modified by the value of $\mu$ as well.} \end{figure}

The retardation effects due to spectral gaps (or rest mass in effective relativistic theories) can be studied by means of (\ref{2.3}) with the expansion (\ref{2.4}). The results are illustrated in figure \ref{fig:6}. Essentially, we obtain an increase of the slopes of caustics as a function of $\mu$, leading to a narrowing of the conical shapes. We infer that signals in binary media are delayed in comparison with homogeneous materials. Also notable is the variation of the pattern periodicity as a function of $\mu$, which manifests in an increase of the distance between maxima and minima in the region inside the spatiotemporal cones. Compare, for example, the first and last panels of figure \ref{fig:6}.

\subsection{Few-body examples}

\begin{figure}[!h] \begin{center} \begin{tabular}{r r} \includegraphics[scale=0.34]{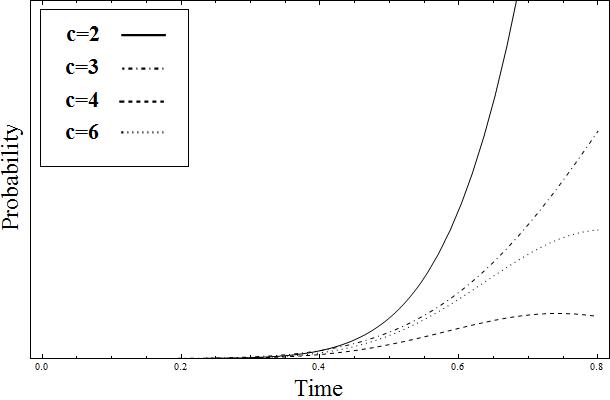} & \includegraphics[scale=0.22]{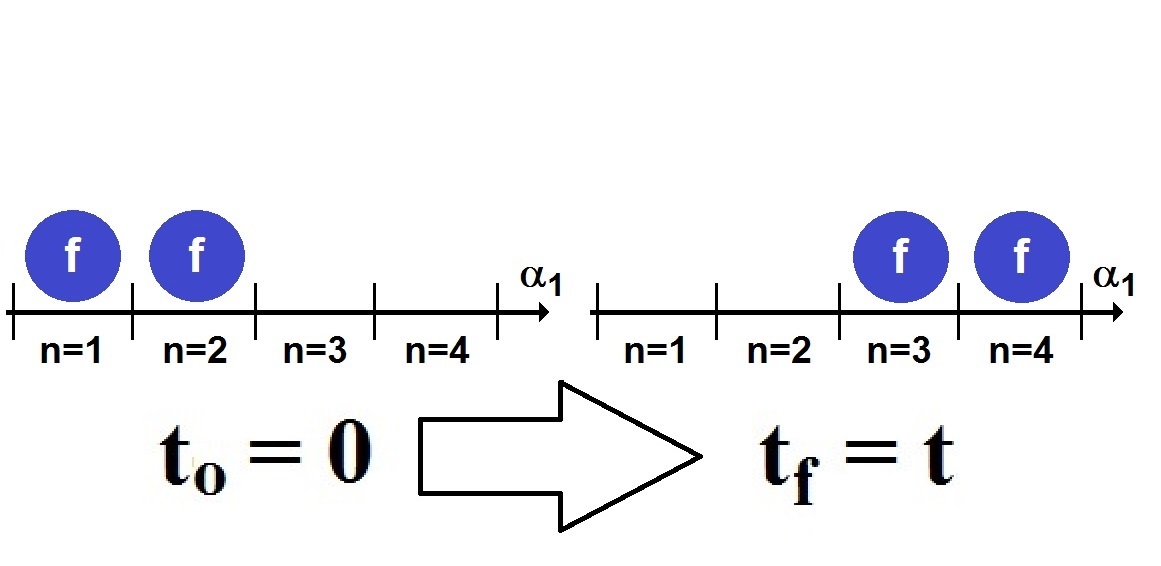} \\    \includegraphics[scale=0.365]{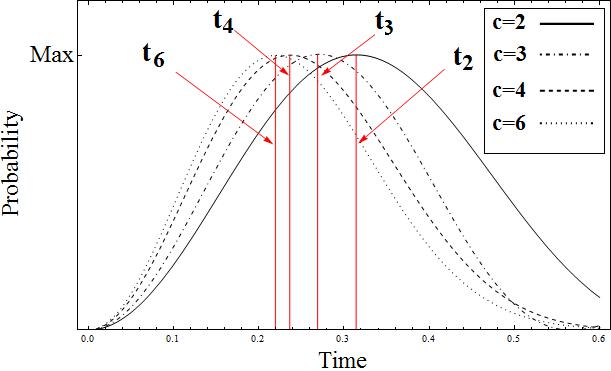} & \includegraphics[scale=0.22]{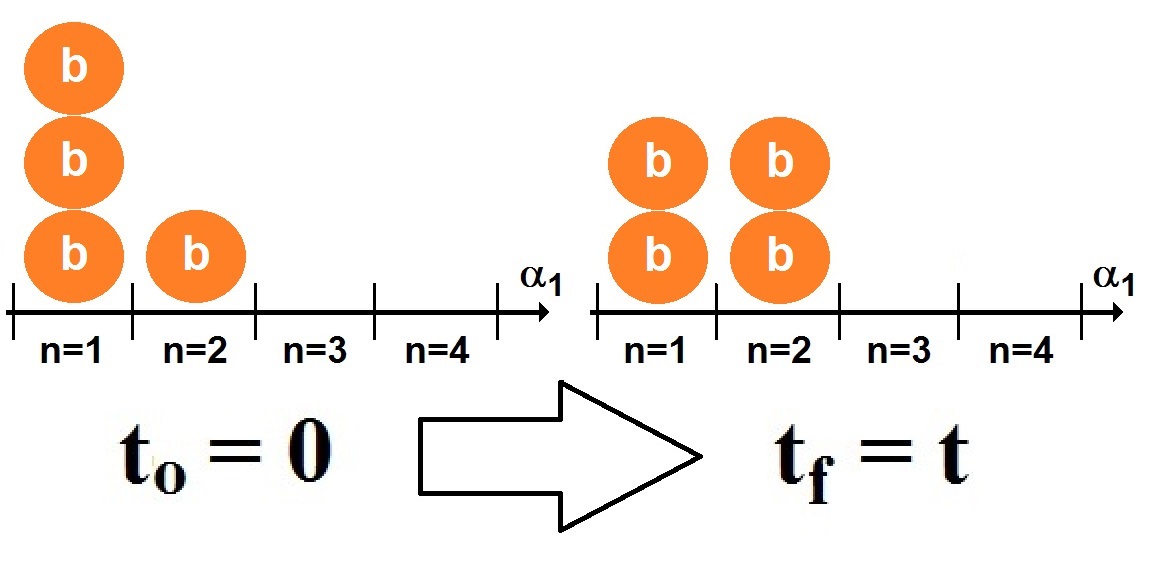} \end{tabular} \end{center} 

\caption{ \label{fig:7} Upper row: Square of the transition amplitudes (left) as functions of time for two fermions (right). The particles migrate from neighboring sites $1,2$ to $3,4$ in the direction of a primitive vector $\bfalpha_i$. Each curve corresponds to a different lattice, indicated by the coordination number $c$. The results show a decrease in the probability with an increasing $c$. Lower row: Square of the transition amplitudes normalized to their maximum value (left) for bosonic process (right). In this computation, four bosons move from occupation numbers $n_1=3,n_2=1$ to $m_1=2,m_2=2$. The results indicate that the propagation occurs faster as the coordination number increases. The times $t_c$ for which a maximum occurs obey the ordering $t_6<t_4<t_3<t_2$.}\end{figure}

With the purpose of understanding the effects of particle statistics in the dynamics, we propose some simple initial distributions on the lattice sites and use (\ref{III.6}) or (\ref{III.7}) for the computation of transition amplitudes. Two examples offer themselves. For fermions, take an intial set of occupation numbers $n_1=1,n_2=1$ corresponding to two particles placed at contiguous sites. Then consider the possibility of finding that the two particles have migrated along a primitive vector in the same juxtaposed configuration, \ie $m_3=1,m_4=1$, see figure \ref{fig:7}. This migratory process without any external stimulus has a non-vanishing probability due to the wavepacket expansions previously studied. The results in the upper row of the figure exhibit a retardation effect in agreement with a Lieb-Robinson bound. In addition, the value of the probability is shown to depend on the coordination number $c$. This can be explained by recognizing that there is more space for expansion from a site to its neigbours as $c$ increases, leading to a decrease in the amplitude. The explicit form of $\< \{m_j\} |U_t|\{n_i\}\>$ for this process has been carried out in the examples of appendix A, formula (\ref{a.22}).

The other example corresponds to bosons. Consider a group of four particles with initial occupation numbers $n_1 = 3,n_2=1$ for contiguous lattice sites. We study the migration of a single boson to an equalized configuration $m_1=2, m_2=2$ at time $t$. The transition amplitude for this process is given explicitly in (\ref{a.13}), while figure \ref{fig:7} shows the results. It should be noted that the various amplitudes for different values of $c$ may differ by orders of magnitude, but we may study the retardation effects in our process by normalizing each amplitude to its maximum as a function of time. Thus, the position of each peak in the normalized signal corresponds to a time of arrival of the bosons to their final configuration. The results change with the value of $c$, as shown by the values of $t_c$ in the plots (lower row of figure \ref{fig:7}). The maximum probability is reached at a time that increases with a {\it decrease\ }of the coordination number, leading to an ordering $t_2 > t_3 > t_4 > t_6$. In conclusion, the signal arrives faster if the lattice is more connected.

%

\section{The Dirac limit viewed in space and time} \label{s5}

A simple way to obtain Dirac equations in hexagonal lattices and dimeric chains entails the direct use of a Fourier transform starting from expansions around conical points in momentum space \cite{semenoff}. It is seldom discussed, however, how the sites in a crystal conspire to form a light-cone in Minkowski space. A rather intuitive approach to the problem comes from the computation of caustics in the spectral decomposition of the propagators. In a very good approximation, one is able to extract light-cones on the lattice from the stationary path of the integral in (\ref{V.5})

\bea
\nabla_{\v k} \left[ \v k\cdot (\v A - \v A') - \lambda_k t \right] = 0 \quad \mbox{for certain} \quad \v k = \v k^* 
\label{d0}
\eea 
where $\lambda_k$ is defined as in equation (\ref{IV.7}). The previous ray equation defines the position of the effective cones at constant phase (caustics):

\bea
\v k^*\cdot (\v A - \v A') - t \lambda_{k^*} = 2\pi n, \quad n \in \v Z.
\label{d0}
\eea 
Although this procedure leads directly to the desired results, it is not clear from the outset how to take the continuous limit in terms of couplings; it is our goal to explore this process. In this respect we should note that continuous limits in lattices can be obtained in a variety of ways, which may lead to relativistic as well as non-relativistic systems: see e.g. \cite{5, yellin}, where restrictions have been imposed specifically to yield the gaussian propagators typical of non-relativistic theories. In the following we show that other ways of specifying continuous limits lead naturally to effective relativistic kernels for hexagonal and linear lattices. We also investigate the emergence of Dirac operators in such limits.

\subsection{Effective Dirac matrices and operators}

Let us start with the algebraic structure. The traditional form of the Dirac propagator entails a definition of matrices that can be transformed covariantly under the Lorentz group. A simple way to obtain them in a crystal is provided by the propagator equation itself; inserting $\v 1 = \tau_3 \tau_3$ between operators we have

\bea
\fl \left\{ H-E_0 -i  \frac{\partial}{\partial t} \right\} \left\{ \tau_3 \tau_3 \right\} \left\{ H-E_0 +i \frac{\partial}{\partial t} \right\} K_S(\v A, \v A';t) = -i \delta_{\v A, \v A'} \delta(t).
\label{d1}
\eea
We restore the coupling constant $\Delta$ in $H$ and substitute (\ref{IV.5}) in (\ref{d1}), leading to the propagator equation
\bea
\fl \left\{ \Delta(\tau_{+}\tau_3 p_{-} + \tau_{-}\tau_3 p_{+})-i \tau_3 \frac{\partial}{\partial t} + \mu \right\} \left\{  \Delta(\tau_{3}\tau_+ p_{-} + \tau_{3}\tau_- p_{+}) + i \tau_3 \frac{\partial}{\partial t} + \mu \right\} K_S(\v A, \v A';t) \nonumber\\ = -i \delta_{\v A, \v A'} \delta(t).
\label{d2}
\eea
Therefore, the Dirac matrices for the lattice can be defined in the form

\bea
\gamma_1 = (\tau_+ + \tau_-)\tau_3 , \qquad \gamma_2= i(\tau_+ - \tau_-)\tau_3, \qquad \gamma_0= \tau_3
\label{d3}
\eea
satisfying the Clifford algebra associated to $2+1$ Minkowski space with metric $\mbox{diag}(+,-,-)$. The cartesian momenta operators $\pi_{1}, \pi_{2}$ must be given then by

\bea
\pi_1 = \frac{1}{2} (p_+ + p_-), \quad \pi_2 = \frac{1}{2i} (p_+ - p_-), \quad \pi_0 = i \frac{\partial}{\partial t}
\label{d4}
\eea
leading finally to

\bea
\left\{ \Delta \gamma^{\nu} \pi_{\nu} - \mu \right\} \left\{ \Delta \gamma^{\nu} \pi_{\nu} + \mu \right\} K_S(\v A, \v A';t)  = i \delta_{\v A, \v A'} \delta(t).
\label{d5}
\eea
In the case of dimeric chains, we can be even more explicit and write

\bea
\gamma_0 = \sum_{n \, \rm{even}}  \left( | n\>\<n|- | n+1\>\<n+1|\right) \nonumber \\
\gamma_1 = \sum_{n \, \rm{even}}  \left( | n+1\>\<n|- | n\>\<n+1|\right) \nonumber \\
\gamma_2 = i\sum_{n \, \rm{even}}  \left( | n+1\>\<n| + | n\>\<n+1|\right).
\label{d6}
\eea
Now that the covariant Dirac matrices have been found, we proceed to obtain the true Lorentz covariant momenta in Minkowski space. We shall impose the continuous limit such that $\pi_{\nu}$ reduces to the usual momentum $p_{\nu}$. For this purpose, the translation operators and their expansion around conical points $\v k_0$ become useful. In the case of the hexagonal lattice, we have primitive vectors $\v b_i$ (see section \ref{hexagon}) and the nearest-neighbor expansion 

\bea
p_{\pm} &=&\sum_{i=1,2,3} \exp \left[\pm i \v p \cdot \v (\v b_i-\v b_1)  \right] \nonumber \\ &=&  \sum_{i=1,2,3} e^{\pm i\v k_0 \cdot (\v b_i- \v b_1)}\exp \left[\pm i \Delta (\v p-\v k_0) \cdot (\v b_i-\v b_1) / \Delta \right]. 
\label{d6.1}
\eea
Rescaling the lattice dimensions such that $\Delta \v p \mapsto \v p$ and Taylor-expanding in $1/\Delta$ we have
\bea
\pi_1 &=& \frac{3}{2\Delta} (\Delta \v k_0 - \v p) \cdot \left[\sum_{i=1,2,3} \sin \left( \v k_0 \cdot \v b_i \right) \v b_i \right] + O(1/\Delta) \nonumber \\
 &\approx& \frac{3}{2\Delta} \left(p_1 -  \frac{4 \pi \Delta}{3 \sqrt{3}} \right) \qquad \mbox{for the conical point} \quad \v k_0 = \frac{4 \pi }{3 \sqrt{3}} \v i 
\label{d7}
\eea
and 
\bea
\pi_2 &=& \frac{3}{2\Delta}(\v p - \Delta \v k_0) \cdot \left[\sum_{i=1,2,3} \cos \left( \v k_0 \cdot \v b_i \right) \v b_i \right] + O(1/\Delta) \nonumber \\
 &\approx& \frac{3}{2\Delta} p_2 \qquad \mbox{for the conical point} \quad \v k_0 = \frac{4 \pi}{3 \sqrt{3}} \v i. 
\label{d8}
\eea
In this way, we are allowed to take the continuous limit by letting $\Delta \rightarrow \infty$ and fixing $\v A / \Delta \rightarrow \v x, \v A' / \Delta \rightarrow \v x'$. With this we show that the inverse of the coupling plays the role of the lattice spacing. This is in sharp contrast with the non-relativistic continuous limit, where the lattice spacing is $\sqrt{\Delta}$, as shown in (\ref{1.4}) using Meissel expansions \cite{1}. Finally, we can always gauge away the conical point with factors $e^{-i\Delta \v k_0 \cdot \v x}$ and obtain the limit of the operator

\bea
\left\{ E_0 - H + i\frac{\partial}{\partial t} \right\}\tau_3 \longrightarrow e^{i\Delta \v k_0 \cdot \v x} \left\{ \gamma^{\nu} p_{\nu} - \mu \right\} e^{-i \Delta \v k_0 \cdot \v x},
\label{d9}
\eea
with an effective speed of light $3/2$ \footnote{In (\ref{d7}) and (\ref{d8}) an overall factor $3/2$ appears due our choice $(\v b)^2=1$, whereas other works \cite{semenoff} use a normalization of the primitive vector  $(\v b)^2=3$. For similar reasons we obtain a factor of 2 in (\ref{d10}). See sections \ref{dirac} and \ref{dirac2}.}. In the case of dimeric chains, a similar treatment is possible. There is only one conical point given by $k=\pi/2$ (we are working in the reduced interval $k \in [0,\pi]$). We replace the primitive vectors $\bfalpha_1 \mapsto -1$, $\bfalpha_2 \mapsto +1$ and obtain
\bea
\pi_1 \approx \frac{2}{\Delta} \left(p -  \frac{\pi \Delta}{2} \right), \quad \pi_2 \approx 0.
\label{d10}
\eea
leading to final results that are similar to (\ref{d9}), but restricted to one dimension in view of $\pi_2 \rightarrow 0$. The gauge factor in this case is $\exp \left(-i \pi \Delta (x-x')/2 \right) = e^{-i(n-n')\pi/2}=i^{n'-n}$.

The specific form of probabilities in space-time and their transition to relativistic cones are also illustrative. This point can be discussed better by analyzing directly the scalar propagators $K_S$ in spectral form, as we show in the following.

\subsection{The Klein-Gordon limit of scalar propagators}

Now that we have found the right limit in terms of the coupling, we can obtain the Klein-Gordon propagator straightforwardly. We must remember that our final result should {\it not\ }coincide with the usual Feynman propagators obtained by the introduction of Wick's angle. Instead, we are expecting a {\it retarded\ }Klein-Gordon propagator which shall vanish identically for points outside of the light-cones. Let us start with

\bea
K_S = \frac{1}{\omega} \int_{\Omega_B} d \v k \exp \left[i \v k \cdot (\v A - \v A') \right] \frac{e^{-i\lambda_k t} - e^{i\lambda_k t}}{2 \lambda_k}.
\label{d11}
\eea
We treat the time-dependent exponentials in the previous expression individually and perform the change of variables $\v q = \Delta (\v k - \v k_0)$, leading to

\bea
 \fl \frac{1}{\Delta^2 \omega} \int_{\Omega} d \v q \frac{\exp \left[i \v q \cdot (\frac{\v A - \v A'}{\Delta}) \pm i \lambda_{q/\Delta} t  \right] }{2 \lambda_{q/ \Delta}} \rightarrow dA \int_{ \v R^2} d \v q \frac{\exp \left[i \v q \cdot (\v x - \v x') \pm i \sqrt{q^2 + \mu^2} t  \right] }{2 \sqrt{q^2 + \mu^2}}  \nonumber \\
\label{d12}
\eea
where $\Omega$ denotes the transformation of $\Omega_B$. The r.h.s. of (\ref{d12}) is known to be a Bessel function \cite{sym, zhang}. Note that the area differential has been defined as $dA \equiv 1/(\Delta^2 \omega)$ and its appearence is justified by a transition from Riemann sums to integrals in the computation of wavefunctions. Finally, the limit of $K_S$ contains the contribution from both signs of the energy in the form

\bea
\fl K_S(\v A, \v A' ; t) \rightarrow dA \int_{ \v R^2} d \v q \frac{ e^{i \v q \cdot (\v x - \v x') - i \sqrt{q^2 + \mu^2} t }- e^{i \v q \cdot (\v x - \v x') + i \sqrt{q^2 + \mu^2} t }}{2 \sqrt{q^2 + \mu^2}}\equiv dA \times K_{\rm{KG}}^{\rm{ret}} (\v x- \v x', t) \nonumber \\
\label{d13}
\eea
with $K_{\rm{KG}}^{\rm{ret}}$ the retarded Klein-Gordon propagator. The kernel vanishes if the events  $(t,\v x')$ and $(0, \v x)$ cannot be connected by a light-cone. Another remark is at hand: when replacing the kernel (\ref{d13}) in (\ref{d5}) it should be noted that the limit of the delta on the r.h.s. of (\ref{d5}) is given by $\delta_{\v A, \v A}\rightarrow dA \times \delta(\v x-\v x')$ and this allows to recover the usual propagator equation

\bea
 \fl \left\{ \gamma^{\nu} p_{\nu} - \mu \right\} \left\{ \gamma^{\nu} p_{\nu} + \mu \right\} \left\{ e^{-i\Delta \v k_0 \cdot (\v x- \v x')} K_{\rm{KG}}^{\rm{ret}}(\v x- \v x',t) \right\} = i \delta(\v x - \v x') \delta(t) \v 1_{2\times 2},
\label{d14}
\eea
where $\v 1_{2\times 2}$ is the identity in spinor space. The full propagator of our problem has therefore the retarded Dirac propagator as its continuous limit

\bea
  K_{\rm{Dirac}}^{\rm{ret}}(\v x- \v x',t)  = \left\{ \gamma^{\nu} p_{\nu} + \mu \right\} e^{-i\Delta \v k_0 \cdot (\v x- \v x')} K_{\rm{KG}}^{\rm{ret}}(\v x- \v x',t). 
\label{d15}
\eea
This procedure can be repeated for the other (inequivalent) conical points in the reciprocal honeycomb lattice.

\subsection{The explicit case of a Weyl propagator}

It is rewarding to work out the case of the monomeric chain without the direct introduction of Dirac operators. This shall clarify the appearance of $*$-spin in terms of the parity of lattice sites. Moreover, this example shall exhibit that the {\it Weyl neutrino\ }is contained as a particular limit of the homogeneous chain. We have

\bea
K(n,m;t) = \frac{1}{2\pi} \int_{0}^{2\pi} dk \, e^{ik(n-m)-i2\Delta t \cos k}.
\label{d16}
\eea
The integral can be expressed in two parts by translating the origin of $k$ to the points of maximal slope of the energy, \ie $q = (k-\pi/2)/\Delta$ and $q = (k-3\pi/2)/\Delta$:

\bea
\fl K(n,m;t) = \frac{1}{2\pi\Delta} \int_{-\frac{\pi \Delta}{2}}^{\frac{\pi \Delta}{2}} dq \, \left[ i^{n-m}e^{iq(\frac{n-m}{\Delta})-i2\Delta t \sin \left(\frac{q}{\Delta} \right)} + i^{m-n} e^{iq(\frac{n-m}{\Delta})+i2\Delta t \sin \left(\frac{q}{\Delta} \right)} \right].
\label{d17}
\eea
By setting $dx \equiv 1/\Delta$ and $x-x' \equiv (n-m)/\Delta$, the limit $\Delta \rightarrow \infty$ can be computed now in the form

\bea
K(n,m;t) &\rightarrow& \frac{dx}{2\pi} \int_{-\infty}^{\infty} dq \, \left[ i^{n-m}e^{iq(x-x')-i2 q t} + i^{m-n} e^{iq(x-x')+i2 q t} \right]  \nonumber \\
&=& \left[dx \right]  i^{m-n}\left[ (-)^{n-m} \delta(x-x'-2t) +  \delta(x-x'+2t) \right].
\label{d18}
\eea
Here it is important to remember our boundary condition at the origin of time by restoring the factor $\theta(t)$ in the full propagator. This, together with the different parities of $n$ and $m$, exhibits the spinorial structure of the propagator (\ref{d18}) as follows: Let $\mbox{lim}_{\Delta \rightarrow \infty} i^{n-m}K(n,m;t) =  dx \, K(x,x';t)$ and denote by  $\mbox{sg}$ the signum function, defined such that $\mbox{sg}(0) \equiv 0$. Since the four cases $n$ even -- $m$ even, $n$ odd -- $m$ even, $n$ even -- $m$ odd and $n$ odd -- $m$ odd can be cast into a $2\times 2$ matrix, we may write 

\bea
\fl K(x,x';t)=  \left(\begin{array}{cc} \delta(2t-|x-x'|) &  \mbox{sg}(x'-x) \delta(2t-|x-x'|) \\  \mbox{sg}(x'-x) \delta(2t-|x-x'|) & \delta(2t-|x-x'|) \end{array} \right).
\label{d19}
\eea
This expression has a close ressemblance with the kernel of the Weyl equation in $1+1$ dimensions, \ie in terms of the Pauli matrix $\tau_1$ we have

\bea
K(x,x';t)= \left\{  \frac{1}{2} \frac{\partial}{\partial t} +\tau_1 \frac{\partial}{\partial x} \right\} \theta(2t - |x-x'| ).
\label{d20}
\eea
This result can be reproduced at the level of the equation as well, for $K(n,m;t)$ absorbs the gauge factor $i^{m-n}$ found in (\ref{d10}) and the equation of the kernel becomes \bea (2 \partial_x - \partial_t)K(x,x';t) =\delta(x-x')\delta(t). \eea

\section{Summary of hamiltonians and propagators} \label{summary}

Here we present the propagators that can be computed or estimated with our previously described techniques. We present them in increasing degree of complexity, starting from linear chains with and without spectral gaps and finishing with hexagonal propagators. We include tables \ref{tab:1} and \ref{tab:2} as a quick guide to the results.

\Table{\label{tab:1} Hamiltonians and propagators.}
\br
Lattice&Symbol&Hamiltonian&Propagator& Auxiliary kernel  \\
\mr
Monomer chain&$\circ-$&(\ref{1.1})&(\ref{1.3})& $--$ \\
Dimeric chain&$\circ-\bullet$&(\ref{2.1})&(\ref{2.3})$^{\rm c}$&(\ref{2.3.1})$^{\rm a}$, (\ref{2.4})$^{\rm b}$\\
Square& $\square$ & (\ref{3.1})  & (\ref{3.3}), (\ref{3.4})$^{\rm c}$ & $--$ \\
Triangular& $\triangledown$ & (\ref{4.1})  & (\ref{4.2}) & $--$ \\
Hexagonal& $\circledast$ & (\ref{5.1})  & (\ref{5.3})$^{\rm c}$ &  (\ref{5.4})$^{\rm a}$, (\ref{5.5})$^{\rm b}$ \\
\br
\end{tabular}
\item[] $^{\rm a}$ Operational form.
\item[] $^{\rm b}$ Bessel expansion.
\item[] $^{\rm c}$ Two species.
\end{indented}
\end{table}

\Table{\label{tab:2} The limits of propagators.}
\br
&& \centre{2}{Continuous} & \centre{2}{Gap-to-coupling ratio}\\
&& \crule{2} & \crule{2} \\
Lattice&Symbol&Relativistic&Non-relativistic& Strong & Weak  \\
\mr
Monomer chain&$\circ-$&(\ref{weyl2})&(\ref{1.4})&$--$&$--$ \\
Dimeric chain&$\circ-\bullet$&(\ref{2.8}), (\ref{2.9})& $--^{\rm d}$&(\ref{2.7})&(\ref{2.6})\\
Hexagonal& $\circledast$ &(\ref{5.7}), (\ref{5.8}) &  $--^{\rm d}$ & (\ref{5.6}) & (\ref{5.5}) \\
\br
\end{tabular}
\item[] $^{\rm d}$A limit of this type can be found from relativistic expressions by taking $\mu \gg 1$. 
\end{indented}
\end{table}

\subsection{Homogeneous chain}
We denote the fundamental cell by $\circ-$
\subsubsection{Hamiltonian}

\bea
H_{\circ-} = \Delta \sum_{n \in \mathbf{Z}} \{ |n \> \<n+1 | + |n+1 \> \<n | \}    
\label{1.1}
\eea
For any function $f$ defined on the integer lattice we have

\bea
\hat H_{\circ-}f_n = \Delta (f_{n+1}+ f_{n-1})
\label{1.2}
\eea

\subsubsection{Propagator}

\bea
K_{\circ-}(n,m;t) = i^{m-n} J_{n-m}(2\Delta t) 
\label{1.3}
\eea
where $J$ is a Bessel function of the first kind. For a similar result in discrete space see \cite{5}.

\subsubsection{Limit 1: Continuous variables} Let $K_{\rm{Free}}$ denote the propagator of a non-relativistic free particle \cite{3.2}. Take $\Delta$ as the inverse square of the lattice spacing, such that $\Delta \rightarrow \infty$. Set $n,m \rightarrow \infty$, but keep fixed the quotient $(n-m)/\sqrt{\Delta} \rightarrow x-x'$, showing the transition from discrete points $n,m$ to continuous variables $x,x'$. Then the following holds 

\bea
 K_{\circ-}(n,m;t) &\rightarrow&  \left[\Delta^{-1/2}\right] e^{-i2t\Delta} \sqrt{\frac{i}{\pi t}} \exp \left[  \frac{ (x-x')^2}{it }  \right] \nonumber \\
&=&  \left[dx' \right]  e^{-i2t\Delta}   K_{\rm{Free}}(x,x';-2t)  
\label{1.4}
\eea
where the differential $dx' \equiv \Delta^{-1/2}$ shows the transition from sums over $m$ to integrals over $x'$. This limit was studied in \cite{5} and it can be computed by taking one stationary point in the Bessel integral \cite{1}.

\subsubsection{Limit 2: Weyl propagator} Let  $\tau_i$ be Pauli matrices and $K_{\rm{Weyl}}$ the $2\times 2$ kernel

\bea
K_{\rm{Weyl}}(x,x';t)=  \left\{ i \tau_2 \frac{\partial}{\partial x} + \tau_3 \frac{\partial}{\partial t} \right\} \theta(t-|x-x'|) \quad t>0
\label{weyl0}
\eea
satisfying the equation
\bea
\left\{ i \tau_3 \frac{\partial}{\partial t}-\tau_2 \frac{\partial}{\partial x}  \right\} K_{\rm{Weyl}}(x,x';t) =  i \delta(x-x')\delta(t)\v 1_{2\times2},
\label{weyl}
\eea
and denote by $K^{i,j}_{\rm{Weyl}}$ the corresponding matrix elements in spinor space. Set the continuous limit such that $dx \equiv 1/\Delta$ (giving directly the lattice spacing), but keep fixed $(n-m)/\Delta \rightarrow x-x'$. Then $\Delta \rightarrow \infty$ yields the following

\bea
\fl i^{m-n} K_{\circ-}(n,m;t) &\rightarrow& \left[ dx \right] \left[ \delta(x-x'-2t) + (-)^{n-m} \delta(x-x'+2t) \right] \nonumber \\
\fl &=& \left[ dx \right] \cases{ K^{1,1}_{\rm{Weyl}}(x,x';2t) & $n$ and $m$ even \\  K^{1,2}_{\rm{Weyl}}(x,x';2t) & $n$ odd $m$ even \\ K^{2,1}_{\rm{Weyl}}(x,x';2t) & $n$ even and $m$ odd \\ K^{2,2}_{\rm{Weyl}}(x,x';2t) & $n$ and $m$ odd }.
\label{weyl2}
\eea

\subsection{Dimeric chain}
We denote the fundamental cell by $\circ-\bullet$. It is possible to define the primitive vectors as $\pm 1$. 
\subsubsection{Hamiltonian}

\bea
H_{\circ-\bullet} &=& \Delta \sum_{n \in \mathbf{Z}} \{ |n \> \<n+1 | + |n+1 \> \<n | \} + E_1 \sum_{n \, \rm{even}} |n\>\<n| + E_2 \sum_{n \, \rm{odd}} |n\>\<n| \nonumber \\
   &=& H_{\circ-} + E_0+ \mu \left( \sum_{n \, \rm{even}} |n\>\<n| - \sum_{n \, \rm{odd}} |n\>\<n| \right)
\label{2.1}
\eea
with $E_0= \frac{1}{2}(E_1+E_2), \mu=\frac{1}{2}(E_1-E_2)$. The spectral gap $\mu$ represents a rest mass in the Dirac limit. The action of $H_{\circ-\bullet}$ on any $f$ reads

\bea
 \hat H_{\circ-\bullet} f_n =  \left(E_0 +(-1)^n \mu \right) f_n + \Delta \left( f_{n+1} + f_{n-1} \right) 
\label{2.2}
\eea

\subsubsection{Propagator}

The kernel $K_{\circ-\bullet}$ can be put in terms of the homogeneous chain kernel $K_{\circ-}$ in the following form

\bea
\fl K_{\circ-\bullet}(n,m;t) &=&  e^{-iE_0t} \left[\hat H_{\circ-\bullet} -E_0 + i \frac{\partial}{\partial t} \right] G(n,m;t)
\label{2.3}
\eea
\bea
\fl G(n,m;t) &=&  \cases{K_{\circ-}\left(n,m; \frac{i\Delta}{2\mu} \frac{\partial}{\partial \mu} \right) \frac{\sin \left( t\sqrt{\mu^2+2\Delta^2} \right)}{i\sqrt{\mu^2 + 2\Delta^2}} & for $n-m$ even \\ 0 & for $n-m$ odd}
\label{2.3.1}
\eea
where $\hat H_{\circ-\bullet}$ acts on the first index of the function to its right using the rule in (\ref{2.2}). The operator $J_{n-m}\left(\frac{i\Delta^2}{\mu} \frac{\partial}{\partial \mu} \right)$ appearing in (\ref{2.3}) is understood as the ascending series of the Bessel function acting on the function of $\mu$ to its right. The auxiliary scalar kernel $G$  (or $K_S$ as in (\ref{VII.1.2})) can be given in alternate forms as follows.

\subsubsection{Spherical wave expansion}
\bea
\fl G(n,m;t) = \cases{4\pi i \Delta t (-1)^{\frac{|n-m|}{2}+1} \sum_{l=0}^{\infty} j_{l}(\mu_{-} t) j_{l}(\mu_{+}t) \left[ P_{l}^{\frac{|n-m|}{2}}(0) \right]^2&  for $n-m$ even \\ 0 & for $n-m$ odd } \nonumber \\
\label{2.4}
\eea
where $\mu_{\pm} \equiv \frac{1}{2}(\sqrt{4\Delta^2 + \mu^2} \pm \mu)$, $j_l$ is the $l$-th spherical Bessel function of the first kind and $P_l^n(0)$ is the normalized associated Legendre polynomial evaluated at the origin.

\subsubsection{Descending series in $\mu$} Separating the sine of (\ref{2.3}) in exponentials, using formula (\ref{b5}) and neglecting higher order terms, we have

\bea
\fl G(n,m;t) &=& \frac{1}{i\sqrt{\mu^2+2\Delta^2}}\left[\frac{1+(-1)^{n-m}}{4} \right] \left[ e^{-it \sqrt{\mu^2 + 2\Delta^2}}K_{\circ-} \left(n,m; \frac{\Delta t}{ 2 \sqrt{\mu^2+2\Delta^2}} \right) \right. \nonumber \\ \fl &-& \left. e^{it \sqrt{\mu^2 + 2\Delta^2}} K_{\circ-} \left(n,m;  \frac{-\Delta t}{ 2 \sqrt{\mu^2+2\Delta^2}} \right)  \right] + O(\Delta^3/\mu^3). 
\label{2.5}
\eea
The expressions (\ref{2.4}) and (\ref{2.5}) for $G$ lead to two different limits. We establish them as follows.

\subsubsection{ Limit 1: Gapless propagator} In the limit $\mu \rightarrow 0$ we recover the homogeneous chain. The corrections in $\mu / \Delta$ come from a Taylor expansion of (\ref{2.4}). Since the linear term of such an expansion vanishes, the first contribution in $\mu$ comes from the operator $H_{\circ-\bullet}$. We obtain

\bea
\fl K_{\circ-\bullet}(n,m;t) &=&  e^{-iE_0t} \left[\hat H_{\circ-\bullet} -E_0 + i \frac{\partial}{\partial t} \right] \times \nonumber \\ &\times& \left[ 4\pi i \Delta t (-1)^{\frac{|n-m|}{2}+1} \sum_{l=0}^{\infty}\left\{ j_l(\Delta t)\, P_{l}^{\frac{|n-m|}{2}}(0) \right\}^2 + O(\mu^2/\Delta^2)\right] \nonumber \\
\fl &=& e^{-iE_0t}\left\{ (-1)^n \mu \, K_{\circ-}(n,m;t) +  K_{\circ-}(n,m;t) \right\} + O(\mu^2/\Delta^2)
\label{2.6}
\eea

\subsubsection{Limit 2: Strong gap} Here we resort to $G$ in its descending form (\ref{2.5}) and re-expand the radicals $\mu_{\pm}$ in $\Delta / \mu$ to obtain

\bea
\fl K_{\circ-\bullet}(n,m;t) &=&  e^{-iE_0t} \left[\hat H_{\circ-\bullet} -E_0 + i \frac{\partial}{\partial t} \right] \times \nonumber \\
&\times&  \left[\frac{1+(-1)^{n-m}}{4 i \mu} \right] \left[ e^{-it\mu}K_{\circ-} \left(n,m; \frac{\Delta t}{ \mu} \right) - e^{it\mu}K_{\circ-} \left(n,m;  \frac{- \Delta t}{\mu} \right)  \right] \nonumber \\ &+& O(\Delta^3/\mu^3)
\label{2.7}
\eea

\subsubsection{ Limit 3: Dirac propagator in $1+1$ dimensions}  \label{dirac} Let us denote the conical point by $k_0= \pi/2$ and arrange the components of the propagator (\ref{2.3}) in a matrix, according the parities of $n$ and $m$. Under the conditions $\Delta \rightarrow \infty$, $dx \equiv 1/\Delta$, $(n-m)/\Delta \rightarrow x-x'$ one has the limit

\bea
K \rightarrow \left[ dx \right] \left\{ \gamma^{\nu} p_{\nu} + \mu \right\} \left\{ e^{-i\Delta \pi (x-x')/2} K^{\rm{ret}}_{\rm{KG}}(x-x',2t) \right\}, \qquad t>0
\label{2.8}
\eea
where $p_{\nu}= \left(i\frac{\partial}{\partial t}, -2i\frac{\partial}{\partial x} \right)$, $\gamma_{\nu}=(\gamma_0, \gamma_1)$ are Dirac matrices as in (\ref{d6}) and $ K^{\rm{ret}}_{\rm{KG}}$  is the retarded propagator in terms of the modified Bessel function $K_1$ (see for example (6.7.71) in \cite{3.2})

\bea
 \fl K^{\rm{ret}}_{\rm{KG}}(x-x';t) &=&  K_{\rm{KG}}(x-x';t) - K_{\rm{KG}}(x-x';-t) \nonumber \\ 
\fl &=& \cases{\frac{2 \mu |t|}{\pi \sqrt{t^2-(x-x')^2}} \, K_1 \left(i \mu \sqrt{t^2-(x-x')^2} \right)& $t>|x-x'|$ \\ 0 & otherwise}
\label{2.9}
\eea

\subsection{Square lattice} \label{square}
We denote the fundamental cell by $\square$ representing four sites. The lattice is spanned by the primitive vectors $\v i,\v j$. We shall cast our operators as functions of integers $n,m,n',m'$, which are related to the lattice vectors by $\v A= n \v i + m \v j, \v A'= n' \v i + m' \v j$.
\subsubsection{Hamiltonian} The hamiltonian for a homogeneous square lattice with first-neighbor interactions reads 
\bea
\fl H_{\square} = \Delta \sum_{n,m \in \v Z} \{ |n+1,m\>\<n,m|+|n,m\>\<n+1,m| + |n,m+1\>\<n,m|+|n,m\>\<n,m+1| \}. \nonumber \\ 
\label{3.1}
\eea
Since the atomic states for this lattice are direct products, \ie $|n,m\> = |n\>\otimes |m\>$, the hamiltonian can be written in terms of homogeneous chain hamiltonians (\ref{1.1}) along $\v i$ and $\v j$ in the form $H_{\square} = H_{\circ-}\otimes \v 1 + \v 1 \otimes H_{\circ-}  \equiv H_{\circ-}^{\v i} +  H_{\circ-}^{\v j} $.

\subsubsection{Propagator} The hamiltonians $ H_{\circ-}^{\v i}$ and $H_{\circ-}^{\v j}$ act on different spaces, therefore we have the following product 

\bea
K_{\square}(\v A, \v A'; t) =  K_{\circ-}(n,n';t)  K_{\circ-}(m,m';t).
\label{3.3}
\eea
This product can be extended to all homogeneous cubic lattices in arbitrary dimensions.
\subsubsection{Generalizations for two species} There are two possible ways to generalize the square lattice to sites of two types. First, we consider rows of different species alternating along $\v i$. Then

\bea
K_{\square}(\v A, \v A'; t) =  K_{\circ-}(n,n';t)  K_{\circ-\bullet}(m,m';t).
\label{3.4}
\eea
The other possibility is to consider two interpenetrating square lattices, each containing one species. The hamiltonian cannot be decomposed in two commuting terms as in the previous case, therefore the propagator does not allow a factorization of the type (\ref{3.4}). In this case one finds a spinorial kernel which can be computed along the lines indicated in section \ref{spectral}.

\subsection{Triangular lattice} \label{triangle}
We denote the fundamental cell by $\triangledown$ representing three sites. For convenience, we adopt the convention $\v a_1 = \frac{1}{2}(\sqrt{3} \v i + 3 \v j) $, $\v a_2 =- \sqrt{3} \v i$ for the primitive vectors generating the lattice and $\v a_3 = -\v a_1 - \v a_2$. The coordination number for this lattice is 6. The vectors connecting one site to its neighbors are denoted by $\bfalpha_i$ and are given in terms of $\pm \v a_i$. Arbitrary lattice vectors are denoted by $\v A = n_1 \v a_1 + n_2 \v a_2, \v A' = n'_1 \v a_1 + n'_2 \v a_2$. The propagators are given in terms of $n_1,n_2,n'_1,n'_2$.
 
\subsubsection{Hamiltonian}

\bea
H_{\triangledown}= \Delta \sum_{\v A, i=1,...,6}\{ |\v A \>\< \v A + \bfalpha_i | + | \v A + \bfalpha_i\>\<  \v A | \} 
\label{4.1}
\eea

\subsubsection{Propagator}

\bea
K_{\triangledown}(\v A, \v A'; t) &=& i^{n'_1+n'_2-n_1-n_2} J^{(+)}_{n_1-n'_1,n_2-n'_2}(2 \Delta t;-i)  \nonumber \\ &=&  I^{(+)}_{n_1-n'_1,n_2-n'_2}(2i \Delta t) 
\label{4.2}
\eea
where $J^{(+)}_{n,m}$ is the two-index Bessel function \cite{1.3} and $I^{(+)}_{n,m}$ is the modified two-index Bessel function. These functions are expressible as series of triple products of Bessel functions of the first kind, therefore the propagator has an alternate form in terms of three coupled kernels $K_{\circ-}$ in directions $\v a_1,\v a_2,\v a_3$:

\bea
K_{\triangledown}(\v A, \v A'; t) = \sum_{s \in \v Z} K_{\circ-}(n_1,n'_1+s;t) K_{\circ-}(n_2,n'_2+s;t) K_{\circ-}(s,0;t)
\label{4.3}
\eea 

\subsection{Hexagonal lattice} \label{hexagon}
We denote the fundamental cell by ${\circledast}$ representing six sites. The coordination number for this lattice is 3. The lattice is decomposable in two triangular sublattices. For convenience, we attach superscripts $+$ and $-$ to functions defined on sublattice A and B respectively. We adopt the convention $\v b_1 = \v j$, $\v b_2 = \frac{1}{2}(\sqrt{3} \v i - \v j)$ for the primitive vectors generating the lattice and $\v b_3 = -\v b_1 - \v b_2$. With this notation we have $\v a_1 = \v b_1 -\v b_3$, $\v a_2 = \v b_3 -\v b_2$. The vectors connecting one site to its neighbors are $\v b_i$.  As before, we have $\v A = n_1 \v a_1 + n_2 \v a_2$, $\v A' = n'_1 \v a_1 + n'_2 \v a_2$. The hexagonal lattice vectors shall be expressed in terms of the triangular vectors $\v A, \v A'$ for type $(+)$ and the addition of $\v b_1$ for sublattice of type $(-)$. The propagator shall be understood as a $2\times2$ matrix, since it must act on spinors. The entries of the propagator shall be functions given in terms of $n_1,n_2,n'_1,n'_2$ in the sense of (\ref{4.2}).  The decomposition of the hexagonal lattice allows to accomodate two species, one in each sublattice  \cite{semenoff}.  
\subsubsection{Hamiltonian}

\bea
\fl H_{ \circledast} &=& \Delta \sum_{\v A, i=1,...,3}\{ |\v A \>\< \v A + \v b_i | + | \v A + \v b_i\>\<  \v A | \} + \sum_{\v A} \{ E_1 | \v A \> \<\v A | +  E_2 | \v A + \v b_1 \> \<\v A + \v b_1 | \} \nonumber \\
\fl &=& \Delta \sum_{\v A, i=1,2,3}\{ |\v A \>\< \v A + \v b_i | + | \v A + \v b_i\>\<  \v A | \} + E_0 \nonumber \\ \fl &+&  \mu  \sum_{\v A} \{ | \v A \> \<\v A | -  | \v A - \v b_1 \> \<\v A + \v b_1 | \},
\label{5.1}
\eea
where $\mu=(E_1-E_2)/2$ and $E_0 = (E_1+E_2)/2$. This operator is related to a triangular hamiltonian in the form $(H_{\circledast}-E_0)^2 = \Delta H_{\triangledown} + \mu^2$. For any spinorial function with components $f^{\pm}$ and triangular lattice variables $n_1,n_2$, we write the action of $H_{\circledast}$ as 

\bea
\fl \hat H_{\circledast} f^{+}_{n_1,n_2} &=& \Delta \left(  f^{-}_{n_1,n_2} +  f^{-}_{n_1+1,n_2} +  f^{-}_{n_1+1,n_2+1} \right) + (E_0 + \mu) f^{+}_{n_1,n_2}  \nonumber \\
\fl \hat H_{\circledast} f^{-}_{n_1,n_2} &=& \Delta \left(  f^{+}_{n_1,n_2} +  f^{+}_{n_1-1,n_2} +  f^{+}_{n_1-1,n_2-1} \right) + (E_0-\mu) f^{-}_{n_1,n_2}
\label{5.2}
\eea
\subsubsection{Propagator} The hexagonal kernel can be written in terms of the triangular propagator (\ref{4.2}). We have the $2 \times 2$ kernel

\bea
\fl K_{\circledast}(\v A, \v A' ; t) =  e^{-iE_0t} \left[\hat H_{\circledast} -E_0 + i \frac{\partial}{\partial t} \right] G_{\triangledown}(\v A,\v A';t)
\label{5.3}
\eea
with the entries of the $2\times 2$ auxiliary $G_{\triangledown}$ given by
\bea
\fl G^{+,+}_{\triangledown}(\v A,\v A';t) &=& G^{-,-}_{\triangledown}(\v A,\v A';t)= K_{\triangledown}\left(\v A,\v A';\frac{i \Delta}{2\mu} \frac{\partial}{\partial \mu} \right) \frac{\sin \left( t\sqrt{\mu^2 + 3 \Delta^2} \right)}{i \sqrt{\mu^2 + 3 \Delta^2}} \nonumber \\
\fl G^{+,-}_{\triangledown}(\v A,\v A';t)&=&G^{-,+}_{\triangledown}(\v A,\v A';t)=0. 
\label{5.4}
\eea
The full kernel (\ref{5.3}) is obtained by the application of the operator in brackets to the object $G_{\triangledown}$ to its right. This is done by following the rule (\ref{5.2}) applied to the indices $n_1,n_2$ of $K_{\triangledown}$ given by (\ref{4.2}) in terms of a two-index Bessel function. Such a function is understood in terms of its ascending series when evaluated at $\frac{i \Delta^2}{\mu} \frac{\partial}{\partial \mu}$. With this general form, some relevant limits can be obtained as follows:

\subsubsection{ Limit 1: Gapless limit (graphene)} The first correction in $\mu$ for $G_{\triangledown}$ is quadratic, whereas for $ K_{\circledast}$ the correction is linear by virtue of (\ref{5.3}). The limit $\mu/\Delta \rightarrow 0$ of the auxiliary kernel is

\bea
\fl G^{+,+}_{\triangledown}(n_1,n_2;n'_1,n'_2;t)=  i^{n'_1 + n'_2 - n_1 - n_2} \sqrt{\frac{8t}{\Delta \pi^3 \sqrt{3}}}\sum_{q=0}^{\infty} C_q^{n_1-n'_1, n_2-n'_2} \left(\frac{i \Delta t}{\sqrt{3}} \right)^q J_{q+\ahalf} (\sqrt{3} \Delta t)  \nonumber \\
\label{5.5}
\eea
with $ G^{-,-}_{\triangledown}= G^{+,+}_{\triangledown}$ and  $ G^{-,+}_{\triangledown}= G^{+,-}_{\triangledown}=0$. The coefficient $ C_q^{n_1-n'_1, n_2-n'_2}$ is given in appendix B, formula (\ref{b.6}). The application of $\hat H_{\circledast}$ on $ G_{\triangledown}$ falls directly on the products $i^{n'_1 + n'_2 - n_1 - n_2} C_q^{n_1-n'_1, n_2-n'_2}$ applying the rules (\ref{5.2}) to indices $n_1,n_2$. 

\subsubsection{ Limit 2: Strong gap limit (boron nitride)} We separate the sine in (\ref{5.4}) in two exponentials and apply to each term the formula (\ref{b5}) to lowest order in $\Delta/\mu$. In terms of the modified two-index Bessel function, we have

\bea
 \fl G^{+,+}_{\triangledown}(n_1,n_2;n'_1,n'_2;t) &=& \frac{1}{i\sqrt{\mu^2+3\Delta^2}} \left[ e^{-it \sqrt{\mu^2 + 3\Delta^2}}  I^{(+)}_{n_1-n'_1,n_2-n'_2}\left(\frac{i\Delta t}{  \sqrt{\mu^2+3\Delta^2}} \right) \right. \nonumber \\ \fl  &-& \left. e^{it \sqrt{\mu^2 + 3\Delta^2}}  I^{(+)}_{n_1-n'_1,n_2-n'_2} \left( \frac{-i\Delta t}{  \sqrt{\mu^2+3\Delta^2}} \right)  \right] + O(\Delta^3/\mu^3).
\label{5.6}
\eea

\subsubsection{ Limit 3: Dirac propagator in $2+1$ dimensions} \label{dirac2}

We repeat the steps leading to (\ref{2.8}) and (\ref{2.9}) for $1+1$ dimensions. However, one has to choose only one conical point from the six available (e.g. $\v k_0 = \frac{4\pi \v i}{3 \sqrt{3}}$) and write

\bea
\fl K \rightarrow \left[ dA\right] \left\{ \gamma^{\nu} p_{\nu} + \mu \right\} \left\{ e^{-i\Delta \v k_0 \cdot (\v x- \v x')} K^{\rm{ret}}_{\rm{KG}}(\v x- \v x',3t/2) \right\}
\label{5.7}
\eea
where $p_{\nu}=( i\frac{\partial}{\partial t}, -\frac{3i}{2}\frac{\partial}{\partial x}, -\frac{3i}{2} \frac{\partial}{\partial y} )$, $\gamma_{\nu} = (\gamma_0, \gamma_1, \gamma_2)$ are the Dirac matrices for this lattice as in (\ref{d3}) and $ K^{\rm{ret}}_{\rm{KG}}$  is the retarded Klein-Gordon propagator given in terms of the modified Bessel function $K_{3/2}$ (as before, see (6.7.71) for $D=2$  in \cite{3.2})

\bea
 \fl K^{\rm{ret}}_{\rm{KG}}(\v x- \v x';t) =  K_{\rm{KG}}( \v x-\v x';t) - K_{\rm{KG}}(\v x-\v x';-t) \nonumber \\ 
\fl = \cases{\left(\frac{(4 i |t|)^{2/3}\mu}{2 \pi i \sqrt{t^2-(\v x-\v x')^2}}\right)^{3/2} \, K_{3/2} \left(i \mu \sqrt{t^2-(\v x- \v x')^2} \right) & $t>|\v x-\v x'|$ \\ 0 & otherwise}
\label{5.8}
\eea

\section{Conclusion} \label{s7}

We have constructed a formalism of propagators in two dimensional lattices with the aim of describing time-dependent processes occurring in crystals. As mentioned in the introduction, the applicability of our results holds for simplified models of condensed matter, e.g. charges moving on long polymeric chains, electrons travelling in graphene and boron nitride or even sufficiently dilute matter waves propagating in optical lattices with a large number of sites. We have given the propagators for particles in every tight-binding {\it regular\ }array as well as the formulae for transition amplitudes in the case of second-quantized many-body dynamics. As a common denominator, we have an explicit form of the Lieb-Robinson bound in terms of Bessel functions and their generalizations. An interesting feature to look at is the coordination number dependence of the results, which exhibits the role of the topology in propagation speeds throughout the arrays. To this end, we have analyzed the example of diffusion of well-localized packets of one or many identical particles for various lattices. Since the corresponding kernels are now available, we may envisage other applications by proposing more general distributions.

From the mathematical point of view, we may give an account of the results by dividing them in two classes: algebraic and functional. In the former case we have shown that a geometric algebra springs from the lattice structure itself, which allows to understand and simplify the results for Green's functions. In the latter case, we have used extensively the Bessel expansions of kernels. We have further identified that previously studied generalizations of Bessel functions --in the context of signal analysis \cite{1.3}-- have naturally appeared in our calculations as the two-index Bessel function gives directly the propagator of the electron in a triangular lattice. 

We should mention that our study in the mathematical direction is not limited by the results presented here: the propagator of a particle subjected to a time-dependent field in one-dimensional discrete space has been computed before on the basis of group theoretical methods and Bessel recurrences \cite{5.1}. We may easily infer that the introduction of constant external fields in two dimensions allows the computation of kernels by similar techniques.

\ack

This work was originated from discussions with Fran\c{c}ois Leyvraz-Waltz and Thomas H Seligman on Dirac limits in graphene; the author is pleased to thank both colleagues. Financial support from PROMEP Project $103.5/12/4367$ is acknowledged. 

\appendix

\section{Transition amplitudes for multiparticle states}

\setcounter{section}{1}

\subsection*{Bosons}

Our purpose here is the calculation of (\ref{III.6}). Let us simplify the notation: we denote the sites by indices $i,j$ instead of $\v A, \v A'$ and express the states with on-site occupation numbers $\{ n_i \}$ at time $t=0$ in the form

\bea
| \{ n_i \} \> \equiv  | \{ n_i \}, t=0 \> = \prod_{i} \frac{ \left[ a^{\dagger}_i (0) \right]^{n_i} }{\sqrt{n_i !}} | \mbox{vac} \>,
\label{a.1}
\eea
where $a^{\dagger}_i (0)$ is the creation operator at site $i$ and $t=0$. Similarly for states $\< \{ m_j \} |$. We are interested in the quantity $\< \{ m_j \} |U_t| \{ n_i \} \>$, which is proportional to the following expression (we omit the factorials in the denominator)

\bea
\fl \< \mbox{vac}, t=0 |  \left\{  \prod_{j}  \left[ a_j (0) \right]^{m_j}  \right\}  U_t  \left\{  \prod_{i}  \left[ a^{\dagger}_i (0) \right]^{n_i}  \right\} | \mbox{vac}, t=0 \> = \nonumber \\
\< \mbox{vac}, t=0 |  \left\{  \prod_{j}  \left[ a_j (0) \right]^{m_j}  \right\} \left\{  \prod_{i}  \left[ a^{\dagger}_i (-t) \right]^{n_i}  \right\} | \mbox{vac}, t \>
\label{a.2}
\eea
where $| \mbox{vac}, t \> = U_t | \mbox{vac}\>$. Now, in order to obtain (\ref{a.2}) explicitly we may use an exponential as a generating function. In terms of derivatives with respect to parameters $x_i, y_j$, we have

\bea
\fl \< \{ m_j \} |U_t| \{ n_i \} \> \propto \prod_i \prod_j \partial_{x_i}^{n_i} \partial_{y_j}^{m_j} \times \nonumber \\
\left[ \< \mbox{vac}, t=0 | \exp\left( \sum_{k} y_k a_k(0) \right) \exp\left( \sum_{l} x_l a^{\dagger}_l(-t) \right)  | \mbox{vac}, t \> \right] \Big|_{x_i=y_j=0}  
\label{a.3}
\eea
The idea of intechanging the exponentials in (\ref{a.3}) for the elimination of irrelevant terms, can be carried out by means of the Baker-Campbell-Hausdorff formula and the commutator of operators at different times:

\bea
\fl \exp\left( \sum_{k} y_k a_k(0) \right) \exp\left( \sum_{l} x_l a^{\dagger}_l(-t) \right) = \nonumber \\ \exp\left( \sum_{i,j} x_i y_j K^*_{ji}(-t) \right)  \exp\left( \sum_{l} x_l a^{\dagger}_l(-t) \right) \exp\left( \sum_{k} y_k a_k(0) \right) 
\label{a.4}
\eea
where $K^*_{ji}(-t) = K_{ji}(t)$ is the propagator as a function of sites $j,i$. Furthermore, from the equations of motion we know that $a_i^{\dagger}(-t) = \sum_j K_{ij}^*(-t) a_j^{\dagger}(0) $, implying

\bea
\fl \< \mbox{vac}, t=0 | \exp\left( \sum_{l} x_l a^{\dagger}_l(-t) \right) &=& \< \mbox{vac}, t=0 | \exp\left( \sum_{l,k} x_l K^*_{lk}(-t) a^{\dagger}_k(0) \right) \nonumber \\ \fl &=& \< \mbox{vac}, t=0 |,
\label{a.5}
\eea
where the first term of the exponential is the only surviving term. Similarly for the kets:
\bea
\fl \exp\left( \sum_{k} y_k a_k(0) \right)| \mbox{vac}, t \> &=& \exp\left( \sum_{k} y_k a_k(0) \right) \exp\left( -it \sum_{i,j} H_{ij} a_i^{\dagger}(0) a_j(0) \right) | \mbox{vac}, t=0 \> \nonumber \\ \fl &=& | \mbox{vac}, t=0 \>. 
\label{a.6}
\eea
Using the two previous relations and the interchangeability formula (\ref{a.4}) in expression (\ref{a.3}), the remaining matrix elements to be computed in (\ref{a.2}) are reduced to

\bea
\fl \< \mbox{vac}, t=0 | \exp\left( \sum_{l} x_l a^{\dagger}_l(-t) \right) \exp\left( \sum_{k} y_k a_k(0) \right)| \mbox{vac}, t \>  &=& \< \mbox{vac}, t=0 | \mbox{vac}, t=0 \>  \nonumber \\ \fl &=& 1. 
\label{a.5}
\eea
Therefore, with the aid of (\ref{a.5}) we reduce (\ref{a.2}) to the expression

\bea
\<\{ m_j \} |U_t| \{ n_i \} \> \propto  \prod_i \prod_j \partial_{x_i}^{n_i} \partial_{y_j}^{m_j}  \exp\left( \sum_{l,k} x_l y_k K_{kl}(t) \right) \Big|_{x_i=y_j=0}.
\label{a.6}
\eea
This relation can be simplified even further by applying the operators $\partial_{x_i}$. We have

\bea
\<\{ m_j \} |U_t| \{ n_i \} \> \propto  \prod_j \partial_{y_j}^{m_j}  \prod_i  \left[\sum_{l}  y_l K_{il}(t) \right]^{n_i}  \Big|_{y_j=0}
\label{a.7}
\eea
and expanding the multinomial that contains the propagators $K_{il}(t)$ leads to

\bea
\fl \<\{ m_j \} |U_t| \{ n_i \} \> \propto  \prod_j  \partial_{y_j}^{m_j}  \prod_i  \sum_{  \{ S\} } n_i ! \prod_{l}\left[ K_{il}(t)^{S_{li}} y_{j}^{S_{li}} \frac{1}{S_{il}!} \right]^{n_i}  \Big|_{y_j=0},
\label{a.8}
\eea
where the summation indices $\{ S\}$ obey the restriction $\sum_{l}S_{li} = n_i$ and relate our calculation with the initial occupation numbers. Now we apply the operators $\partial_{y_j}$ and evaluate at zero, leading to

\bea
\fl \<\{ m_j \} |U_t| \{ n_i \} \> &\propto&  \sum_{  \{ S \} }  \left[  \prod_{i,l} \frac{  n_i ! K_{il}(t)^{S_{li}}   }{  S_{li} !  } \right] \left[ \prod_l \partial_{y_l}^{m_l} 
 y_l^{\sum_i S_{li}} \right]  \Big|_{y_j=0} \nonumber \\ \fl &=& \sum_{  \{ S \} } \prod_{i,l} \frac{n_i ! m_l ! K_{il}(t)^{S_{li}} }{S_{li}!}
\label{a.9}
\eea
which is true as long as the exponent of $y_l$ satisfies $\sum_{i} S_{li}=m_l$; a vanishing result is obtained otherwise. Therefore, the restrictions found on the summation indices $ \{ S\}$ lead naturally to the important result of particle conservation, \ie

\bea
\sum_{ij} S_{ij} = \sum_{j} m_j = \sum_{i} n_i \equiv N.
\label{a.10}
\eea
Restoring the proportionality factors and renaming the index $l \rightarrow j$ yields the final formula

\bea
\<\{ m_j \} |U_t| \{ n_i \} \> = \sum_{  \{ S\} } \prod_{i,j} \frac{ \sqrt{n_i ! m_j !} K_{ij}(t)^{S_{ji}} }{S_{ji}!}.
\label{a.11}
\eea
As an example of the usage of this formula, consider the following two matrices

\bea
\fl S^{(1)} = \begin{array}{ccc} \left( \begin{array}{c} 1 \\ 1 \end{array} \right. & \left. \begin{array}{c} 0 \\ 2 \end{array} \right)  &  \begin{array}{c} n_1=1 \\ n_2 = 3 \end{array} \\
m_1=2 & m_2=2 & N=4 \end{array}, \qquad S^{(2)} = \begin{array}{ccc} \left( \begin{array}{c} 0 \\ 2 \end{array} \right. & \left. \begin{array}{c} 1 \\ 1 \end{array} \right)  &  \begin{array}{c} n_1=1 \\ n_2 = 3 \end{array} \\
m_1=2 & m_2=2 & N=4 \end{array},
\label{a.12}
\eea
where the sums over their rows and their columns are indicated to the right and bottom respectively. These are the only possible matrices satisfying the restrictions for the initial occupation numbers $n_1=1, n_2=3$ at sites $1,2$ and final occupation numbers 
$m_1 = 2, m_2 = 2$ at the same sites. The total number of particles is $N=4$. The transition amplitude is thus given by a sum running on the two possible matrices (\ref{a.12}):

\bea
\fl \<\{ m_j \} |U_t| \{ n_i \} \> &=& \prod_{i,l} \frac{ \sqrt{n_i ! m_l !} K_{il}(t)^{S^{(1)}_{li}} }{S^{(1)}_{li}!} + \prod_{i,l} \frac{ \sqrt{n_i ! m_l !} K_{il}(t)^{S^{(2)}_{li}} }{S^{(2)}_{li}!} \nonumber \\ \fl &=& \sqrt{24} \left[ \frac{1}{2} K_{11}(t)K_{21}(t) K_{22}^{2}(t) + \frac{1}{2} K_{12}(t) K_{21}^2(t) K_{22}(t) \right]  \nonumber \\ \fl &=& \sqrt{6} K_{21}(t) K_{22}(t) \left[  K_{11}(t)K_{22}(t) +  K_{12}(t)K_{21}(t) \right].
\label{a.13}
\eea
As a second example, consider the case in which the occupation numbers of initial and final states coincide, corresponding to an amplitude of permanence of the state. The matrices $S$ must be symmetric and formula (\ref{a.11}) becomes a sum over all symmetric matrices such that the addition of elements in the $n$-th row equals the addition of elements in the $n$-th column. For occupation numbers $m_1=n_1=3$ and $m_2=n_2=2$ there are two possibilities: 

\bea
\fl S^{(1)} = \begin{array}{ccc} \left( \begin{array}{c} 1 \\ 2 \end{array} \right. & \left. \begin{array}{c} 2 \\ 0 \end{array} \right)  &  \begin{array}{c} n_1=3 \\ n_2 = 2 \end{array} \\
m_1=3 & m_2=2 & N=5 \end{array}, \qquad S^{(2)} = \begin{array}{ccc} \left( \begin{array}{c} 3 \\ 0 \end{array} \right. & \left. \begin{array}{c} 0 \\ 2 \end{array} \right)  &  \begin{array}{c} n_1=3 \\ n_2 = 2 \end{array} \\
m_1=3 & m_2=2 & N=5 \end{array},
\label{a.14}
\eea
leading to an amplitude

\bea
\fl \<\{ m_j \} |U_t| \{ n_i \} \> &=& 3! \, 2! \left[ \frac{1}{4} K_{11}(t)K_{12}^{2}(t) K_{21}^{2}(t) + \frac{1}{12} K_{11}^3(t)K_{22}^{2}(t) \right]. 
\label{a.15}
\eea
These examples can be produced by replacing lattice indices $\v A, \v A'$ instead of $i,j$. The results presented in this appendix are valid for general tight-binding lattices, which can be even more complex than the ones treated in this paper.

\subsection*{Fermions}

The calculation of (\ref{III.7}) is considerably simpler than the bosonic case (\ref{III.6}). The occupation numbers only take the values $n_i=0,1$ and $m_j=0,1$. Let us denote the fermionic operators by $f$. We have

\bea
\fl \<\{ n_i \} |U_t| \{ m_j \} \> = \< \mbox{vac}, t=0 |  \left\{  \prod_{i}  \left[ f_i (0) \right]^{n_i}  \right\} \left\{  \prod_{j}  \left[ f^{\dagger}_j (-t) \right]^{m_j}  \right\} | \mbox{vac}, t \>.
\label{a.16}
\eea
Using the equations of motion, we have


\bea
f^{\dagger}_j(-t)^{m_j} = \left[\sum_l K_{jl}^*(-t) f^{\dagger}_l(0) \right]^{m_j} = \left[\sum_l K_{jl}(t) f^{\dagger}_l(0) \right]^{m_j} 
\label{a.17}
\eea
and the products in (\ref{a.16}) have the following possibilities

\bea
\fl \left[f_i (0) \right]^{n_i}  \left[\sum_l K_{jl}(t) f^{\dagger}_j(0) \right]^{m_j}   =\cases{ f_i(0) & if $n_i=1,m_j=0$\\ \sum_l K_{jl}(t) f^{\dagger}_j(0)  & if $n_i=0,m_j=1$
\\ 1 & if $n_i=m_j=0$ \\ \sum_{l} K_{jl}(t) \left[ \delta_{il} - f^{\dagger}_l(0) f_i(0) \right] & if $n_i=m_j=1$      } \nonumber \\
\label{a.18}
\eea
where in the last row we have used the anticommutator at equal times. The only interesting case comes precisely from the fourth possibility, as the others lead to annhilation of the vacua states, either to the left or to the right. Replacing (\ref{a.18}) in (\ref{a.16}) we can see that $f^{\dagger}_j(0)$ destroys the vacuum to its left, \ie for $n_i=m_j=1$ we have

\bea
\fl \< \mbox{vac}, t=0  | \prod_{i,j} \left[ K_{ji}(t) - \sum_{l} K_{jl}(t) f^{\dagger}_l(0) f_i(0) \right] | \mbox{vac}, t  \>  &=& \< \mbox{vac}, t=0  | \prod_{i,j} K_{ji}(t) | \mbox{vac}, t  \> \nonumber \\ \fl &=& \prod_{i,j} K_{ji}(t), 
\label{a.19}
\eea
where we have used again

\bea
\fl \< \mbox{vac}, t=0  | \mbox{vac}, t  \> &=& \< \mbox{vac}, t=0  | \exp \left(-it \sum_{ij}H_{ij} f^{\dagger}_i(0) f_j (0) \right) |\mbox{vac}, t=0  \> \nonumber \\ &=& \< \mbox{vac}, 0  | \mbox{vac}, 0  \> = 1.
\label{a.20}
\eea
Incorporating all cases in (\ref{a.18}), we obtain a quite simple result

\bea
\<\{ n_i \} |U_t| \{ m_j \} \> = \prod_{ij}' K_{ji}(t),
\label{a.21}
\eea
with the product $\prod_{ij}'$ extending over all occupied sites. 

Let us illustrate this with an example. If $N=2$, we may consider an initial configuration with sites $1,2$ occupied by one fermion each, and a final configuration where the fermions have migrated to other sites, say $3$ and $4$. This implies $n_1 =  n_2 = 1$,  $m_3 =  m_4 = 1$ and the rest are zeros. We have 

\bea
\<\{ 1_1,1_2 \} |U_t| \{ 1_3,1_4 \} \> =  K_{13}(t)K_{14}(t)K_{23}(t)K_{24}(t).
\label{a.22}
\eea
As another example, take $N=1$. Our formula gives the probability of migration of a single fermion from site $i$ to $j$. We recover the propagator itself:

\bea
\<\{ 1_i \} |U_t| \{ 1_j \} \> =  K_{ij}(t).
\label{a.23}
\eea

\section{ Ascending series of the two-index Bessel function}

Here we derive the coefficients $C_{s}^{mn}$ in the expansion (\ref{VIII.1}). There are two possible methods.

\subsection*{Method 1: Triple product expansion} Using the general expansion of the two-index Bessel function of {\it two\ }variables \cite{1.3}

\bea
J_{mn}(x;\xi) = \sum_{l \in \v Z} \xi^l J_{m-l}(x)J_{n-l}(x)J_l(x)
\label{b.1}
\eea
and the ascending series of each Bessel function, we have
\bea
\fl J_{mn}(x;-i) &=& \sum_{l \in \v Z} i^{-l} J_{m-l}(x)J_{n-l}(x)J_l(x) \nonumber \\
\fl &=& \sum_{l \in \v Z} i^{-l} \sum_{k_1, k_2, k_3 \in \v N_0}  \left( \frac{x}{2} \right)^{2(k_1+k_2+k_3)+ m+n-l}(-1)^{k_1+k_2+k_3} \times \nonumber \\ &\times&  \frac{1}{k_1 ! k_2 ! k_3 ! (k_1+m-l)!(k_2+n-l)!(k_3+l)!}.
\label{b.2}
\eea
The following redefinition of indices is convenient

\bea
 s \equiv 2(k_1+k_2+k_3)+ m+n-l, \quad n_1 \equiv k_1, \quad n_2 \equiv k_2, \quad n_3 \equiv k_3, \nonumber \\
 n_4 \equiv k_1+m-l, \quad n_5 \equiv k_2+n-l, \quad n_6 \equiv k_3+l,
\label{b.3}
\eea
with the obvious restrictions
\bea
\sum_{i=1}^{6} n_i = s, \quad n_6-n_3-n_1+n_4 = m, \quad n_6-n_3-n_2+n_5= n.
\label{b.4}
\eea
The relations (\ref{b.3}) and (\ref{b.4}) turn (\ref{b.2}) into
\bea
J_{mn}(x;-i) = \sum_{s = 0}^{\infty} \frac{\left( \frac{x}{2} \right)^{s} i^{s-m-n}}{s!} \sum_{\{ n \}}  \left( \begin{array}{c} s \\ \{ n \}  \end{array} \right)  (-1)^{n_1+n_2+n_3},
\label{b.5}
\eea
where $\left( \begin{array}{c} s \\ \{ n \}  \end{array} \right)$ denotes a multinomial coefficient of six indices and $\sum_{\{ n \}}$ is the sum with restrictions (\ref{b.4}). More precisely, the sum can be shown to run over the three indices $n_1, n_2, n_3$ by eliminating $n_4, n_5, n_6$ using the relations

\bea
n_4 = s -n- 2 n_3 -2 n_2 -n_1,
\label{b.5.1}
\eea
\bea
 n_5 = s-2(n_1+n_3) - n_2 - m,
\label{b.5.2}
\eea
\bea
n_6 = 3 n_3 +2(n_1+n_2) + n+m -s.
\label{b.5.3}
\eea
Finally, the expression for the general coefficient can be read from (\ref{b.5}) as 
\bea
C_s^{mn} = \sum_{n_1, n_2, n_3} \frac{i^{s-m-n} (-1)^{n_1+n_2+n_3}}{2^s s!} \left( \begin{array}{c} s \\ \{ n \}  \end{array} \right)
\label{b.6}
\eea
\subsection*{Method 2: Generating function} We start with the expansion

\bea
\fl \exp \left\{ i x \left[ \sin \varphi + \sin \theta - \cos(\varphi+ \theta) \right] \right\} &=& \sum_{n,m \in \v Z} e^{i(m\varphi + n\theta )} J_{mn}(x;-i) \nonumber \\
\fl &=&  \sum_{s=0}^{\infty} \sum_{n,m \in \v Z} e^{i(m\varphi + n\theta )}  C_{s}^{mn} x^s \nonumber \\
\fl &=&  \sum_{s=0}^{\infty} \frac{(ix)^s}{s!}  \left[ \sin \varphi + \sin \theta - \cos(\varphi+ \theta) \right]^s,
\label{b.7}
\eea
which implies

\bea
 \sum_{n,m \in \v Z} e^{i(m\varphi + n\theta )}  C_{s}^{mn} =  \frac{i^s}{s!}  \left[ \sin \varphi + \sin \theta - \cos(\varphi+ \theta) \right]^s.
\label{b.8}
\eea
By simple Fourier integration, we obtain

\bea
\fl C_{s}^{mn} = \frac{i^s}{2 \pi s!} \int_{0}^{2\pi} d\varphi \int_{0}^{2\pi} d\theta  e^{-i(m\varphi + n\theta )}  \left[ \sin \varphi + \sin \theta - \cos(\varphi+ \theta) \right]^s.
\label{b.9}
\eea
The result (\ref{b.6}) follows by expanding the multinomial inside the integral in terms of $e^{-i\varphi}$, $e^{-i\theta}$ and using the orthogonality of Fourier functions. Alternatively, we may define  $z_1=e^{i\varphi}$, $z_2=e^{i\theta}$ and evaluate $C_s^{mn}$ by the method of residues:

\bea
\fl C_{s}^{mn} = -\frac{2^{-s}}{2 \pi s!} \oint dz_1 \oint dz_2 \, z_1^{-m-1}z_2^{-n-1}  \left[ z_1-z_1^{-1} +z_2-z_2^{-1} -i z_1z_2-i(z_1z_2)^{-1} \right]^s
\label{b.10}
\eea
where the contours enclose the origin of the complex planes of $z_1$ and $z_2$. Expanding the multinomial and recognizing that the only contribution comes from simple poles, we get (\ref{b.6}) with the restrictions (\ref{b.5.1}), (\ref{b.5.2}), (\ref{b.5.3}).

\end{document}